\documentclass[fleqn,useAMS,usenatbib,english]{mnras}
\usepackage[T1]{fontenc}
\usepackage{ae,aecompl}
\usepackage[latin9]{inputenc}
\usepackage{verbatim}
\usepackage{amstext}
\usepackage{amssymb}
\usepackage{graphicx}
\usepackage{amsmath}
\usepackage{esint}
\usepackage{times}
\usepackage[switch]{lineno}
\makeatletter

 \@ifundefined{textcolor}{}
 {%
   \definecolor{BLACK}{gray}{0}
   \definecolor{WHITE}{gray}{1}
   \definecolor{RED}{rgb}{1,0,0}
   \definecolor{GREEN}{rgb}{0,1,0}
   \definecolor{BLUE}{rgb}{0,0,1}
   \definecolor{CYAN}{cmyk}{1,0,0,0}
   \definecolor{MAGENTA}{cmyk}{0,1,0,0}
   \definecolor{YELLOW}{cmyk}{0,0,1,0}
 }

\usepackage{hyperref}
\hypersetup{
    colorlinks,%
    citecolor=blue,%
    filecolor=blue,%
    linkcolor=blue,%
    urlcolor=blue
}
\usepackage[all]{hypcap}

\providecommand{\eprintc}[1]{\href{http://arxiv.org/abs/#1}{#1}}
\providecommand{\adsurl}[1]{\href{#1}{ADS}}
\providecommand{\ISBN}[1]{\href{http://cosmologist.info/ISBN/#1}{ISBN: #1}}

\makeatother

\usepackage{babel}
\global\long\def\av#1{\left\langle #1\right\rangle }
 \global\long\def\bx{{\mathbf{x}}}
\global\long\def\d{\mathrm{d}}
\global\long\def\hMpc{h^{-1}\mathrm{Mpc}}
\global\long\def\hDens{h^{3}\mathrm{Mpc}^{-3}}
\global\long\def\hVolG{h^{-3}\mathrm{Gpc}^{3}}

\global\long\def\pasj{PASJ}
\global\long\def\aap{AAP}
\global\long\def\prd{PRD}
\global\long\def\pre{PRE}
\global\long\def\apjl{APJL}
\global\long\def\apjs{APJS}

\begin{document}
\label{firstpage}
\pagerange{\pageref{firstpage}--\pageref{lastpage}}

\title[Minkowski Functionals of BOSS Galaxies]{The clustering of galaxies in the SDSS-III Baryon Oscillation Spectroscopic
Survey: higher-order correlations revealed by germ-grain Minkowski Functionals}

\author[A. Wiegand, D. J. Eisenstein]{Alexander Wiegand\thanks{E--mail:jwiegand@cfa.harvard.edu}, Daniel J. Eisenstein\\
Harvard-Smithsonian Center for Astrophysics, 60 Garden St., Cambridge,
MA 02138, USA }

\maketitle

\begin{abstract}
We probe the higher-order clustering of the galaxies in the final
data release (DR12) of the Sloan Digital Sky Survey Baryon Oscillation
Spectroscopic Survey (BOSS) using the method of germ-grain Minkowski
Functionals (MFs). Our sample consists of $410,615$ BOSS galaxies
from the northern Galactic cap in the redshift range $0.450$--$0.595$.
We show the MFs to be sensitive to contributions up to the six-point
correlation function for this data set. We ensure with a custom angular
mask that the results are more independent of boundary effects than
in previous analyses of this type. We extract the higher-order part of the MFs and quantify
the difference to the case without higher-order correlations. The
resulting $\chi^{2}$ value of over $10,000$ for a modest number
of degrees of freedom, $O(200)$, indicates a $100$-sigma deviation
and demonstrates that we have a highly significant signal of the non-Gaussian
contributions to the galaxy distribution. This statistical power can
be useful in testing models with differing higher-order correlations.
Comparing the galaxy data to the QPM and MultiDark-Patchy mocks, we
find that the latter better describes the observed structure. From
an order-by-order decomposition we expect that, for example, already
a reduction of the amplitude of the MD-Patchy mock power spectrum
by 5\% would remove the remaining tension.
\end{abstract}


\section{Introduction}

The analysis of cosmic large-scale structure in the distribution of
galaxies contributed significantly to establish the current cosmological
concordance model. With the increasing degree of perfection of the
standard two-point correlation analysis, the related measurements
reached unprecedented precision%
. However, the available and upcoming galaxy redshift survey data
\citep{2013AJ....145...10D,2016AJ....151...44D,2013arXiv1308.0847L,2011arXiv1110.3193L}
cannot merely be characterized by its two-point correlation properties.
Already the filamentary structure that we see in those surveys indicates
that the higher-order correlations play an important role. There have
been different approaches to use this information for improving our
knowledge about the constituents and evolution of the Universe.

The most direct approach of course is to simply measure the higher-order
correlations directly. This has been done for both simulated and observed
data starting for early catalogs with \citet{1975ApJ...196....1P} and \citet{1978ApJ...221...19F}.
Especially measurements of the three-point function have been standard
for recent surveys \citep{2011ApJ...737...97M,2013MNRAS.432.2654M,2015arXiv151202231S}
and provide useful complimentary information \citep{2015MNRAS.448....9S}.
For higher orders, however, there are computational and conceptual
challenges. On the computational side, increasing the order of the
correlation function measured unleashes the curse of combinatorics.
Measuring all pairs in current surveys is feasible, for all triangles
there are methods to do it \citep{2015arXiv151202231S}, but beyond
that it becomes increasingly hard. And then, from the conceptual side,
even if measuring the full $n$-point function was possible, it would
require modeling and covariances for a large number of data bins.
Already for a modest number of bins %
{} for each parameter of the $n$-point function this is a challenge.
Therefore, it is worthwhile to consider other approaches that contain
information on certain aspects of the higher-order clustering without
referring to the full correlation functions.

Since the beginning of large-scale structure analysis, a large number
of such methods have been used, including moments of counts in cells,
void probability \citep{Stratonovich1963,1979MNRAS.186..145W}, structure
functions on minimal spanning trees, wavelet methods, the genus, etc.
Recently also approaches that attempt to bring the higher-order information
back into the two-point correlation function through non-linear transforms
were used like %
{} a logarithmic transformation \citep{2009ApJ...698L..90N,2011AAS...21823304N,2011ApJ...735...32W}.

The method (re-)considered in this paper is using Minkowski Functionals%
. They also also fall into the category of methods that provide condensed
access to higher-order information. This information is encoded in
the morphology of the density field. Minkowski Functionals access
it by quantifying extended regions of the field through their geometrical
properties like volume and surface (see Sec.~\ref{sub:Minkowski-Functionals}).
For the construction of these extended regions, two different methods
have been widely used in the past\footnote{See \citet{2010arXiv1006.4178A} for a recent alternative.}: 

In the first application of Minkowski Functionals for analyzing large-scale
structure in the galaxy distribution in \citet{1994A&A...288..697M}
(see \citet{1995lssu.conf..156B} for a short motivation and \citet{1996dmu..conf..281S}
for a brief overview), the regions considered were defined by the
union of balls around every galaxy location. We shall use the same
method, referred to as germ-grain model, here and explain the details
in Sec.~\ref{sub:Germ--grain}. After a couple of applications in
the analysis of galaxy and cluster catalogs \citep{1996app..conf...83K,1998A&A...333....1K,2001A&A...373....1K,1997MNRAS.284...73K},
the germ-grain model has been less used in recent years.

The second and most popular form of the application of Minkowski Functionals
works with extended regions that are defined by iso-contours of the
field under investigation. It has been used for the higher-order contributions
to the dark matter overdensity field \citep{1996app..conf..251P,1997ApJ...482L...1S,1998ApJ...495L...5S,1998ApJ...508..551S,1999ApJ...526..568S,2003PASJ...55..911H,2004astro.ph..8428N,2014MNRAS.437.2488B,2013ApJS..209...19C},
for the weak lensing shear field \citep{2012PhRvD..85j3513K,2013PhRvD..88l3002P},
for local morphology of cosmic structure \citep{2014A&A...562A..87E}
and for the morphology of reionization bubbles \citep{2006MNRAS.370.1329G,2016arXiv160202351Y}.
Another very active field of recent application is the study of iso-temperature
contour maps of the cosmic microwave background (CMB) \citep{1998MNRAS.297..355S,2000ApJ...544L..83S,2012MNRAS.425.2187H,2013MNRAS.429.2104D,2013MNRAS.428..551M,2013MNRAS.434.2830M,2014A&A...571A..23P,2014A&A...571A..25P,2016A&A...594A..16P},
where Minkowski Functionals helped to establish stringent bounds on
the non-Gaussianity of the CMB. So the iso-contour Minkowski functionals
are a well established tool in various domains of cosmology.

However, as we shall discuss in Sec.~\ref{sub:Germ--grain}, the
germ-grain model accesses different aspects of the higher-order distribution
and can be better suited for the analysis of galaxy redshift surveys.
We have demonstrated some of these advantages already in the analysis
of the Sloan Digital Sky Survey (SDSS) Data Release seven (DR7) Luminous
Red Galaxies (LRGs) in \citet{2014MNRAS.443..241W}.

In the present paper, we concentrate more explicitly on the higher-order
aspects of the Minkowski Functionals. To this end, we exploit the
analytically known connection to the integrals over higher-order correlation
functions. Using direct measurements of the two-point correlation
function, we can isolate the contribution of higher-order correlation
functions to the Minkowski Functionals. This allows us to quantify
the evidence for non-Gaussianity of the galaxy distribution (see \citet{2012MNRAS.423.3209P}
and \citet{2013MNRAS.435..531C} for related recent studies of non-Gaussianity
with isodensity contour functionals).

The analysis is demonstrated with the final data from the Baryon Oscillation
Spectroscopic Survey (BOSS) \citep{2013AJ....145...10D} that is now
publicly available within the DR12 of SDSS-III. Being the largest
spectroscopic galaxy redshift data set yet observed, it provides an
ideal testbed for the analysis and reveals the strong presence of
higher-order correlations already at redshifts of $\approx0.55$.

We proceed as follows: Sec.~\ref{sec:Minkowski-Functionals-in-the}
recalls the properties of Minkowski Functionals in general and the
germ-grain model in particular. We write out the connection of the
functionals to higher-order correlations explicitly in \ref{sub:DensDef}
and lay out in \ref{sub:Density-dependence}, how knowing the concrete
form allows us to isolate the part of the functionals that is only
dependent on higher-orders. We explain the subtleties of a Gaussian
reference model in the discrete case in \ref{sub:Gauss-Poisson} and
recall the basic properties of our Minkowski code in \ref{sub:SamplingCode}.

In Sec.~\ref{sec:Data} we turn to the description of the SDSS DR12
data used. Sec.~\ref{sub:CMASS} summarizes the data and mocks we
are using. In \ref{sub:Ceometry}, we address issues related to the
imperfections in the data, constructing a custom angular mask that
defines a high quality region. In \ref{sub:Radial} we turn to the
radial distribution of the data and choose the analysis volume and
the radial grid. In \ref{sub:Weights} we finally discuss the effect
of the regions that can not be accounted for by our custom mask.

Sec.~\ref{sec:HigherOrders} then turns to the core of the analysis.
In \ref{sub:MFdensities} we present the Minkowski Functionals of
the mocks and the data and explain how we transform the data to be
able to isolate the higher-order only part. This part is then examined
in \ref{sub:HighSubtract}. In \ref{sub:SeriesCoefficients} we demonstrate
that the contributions to this higher-order part range at least up
to the $6$-point function. To wrap up we discuss in \ref{sub:Voidfraction}
an interesting aspects about the galaxy distribution using the base
Minkowski Functional to study the volume in extreme voids in the observed
field.

We conclude in Sec.~\ref{sec:Conclusion}.

\section{Minkowski Functionals in the germ--grain model}\label{sec:Minkowski-Functionals-in-the}

In this section, we shall provide the basic notions necessary for
understanding the quantities we used in our analysis. We summarize
some key points already covered in the DR7 analysis paper \citep{2014MNRAS.443..241W},
but with an emphasis on the crucial points that are the source for
common misconceptions.

\subsection{Minkowski Functionals\label{sub:Minkowski-Functionals}}

Minkowski Functionals (often called MFs later on) quantify the morphology
of extended bodies. They are defined as scalar functionals that map
the shape of a body to a set of real numbers. A quite surprising result
of integral geometry \citep{UBHD152758} is that any such scalar functional,
defined for bodies in Euclidean $d$-dimensional space, that fulfills
some basic requirements (motion invariance, additivity and conditional
continuity) can be expressed as a linear combination of $d+1$ base
functionals. 

There are different choices for the normalization of the base functionals.
We will use one where the four MFs that exist for $d=3$ are denoted
$V_{0}-V_{3}$ and related to geometrical properties of the body as
follows: 
\begin{equation}
V_{0}=V\;\;;\;\;V_{1}=\frac{S}{6}\;\;;\;\;V_{2}=\frac{H}{3\pi}\;\;;\;\;V_{3}=\chi\;.\label{eq:MinkRelGeo}
\end{equation}
Here, $V$ is the volume of the body, $S$ is its surface area, $H$
the integral mean curvature of the surface, and $\chi$ the Euler
characteristic (the integral Gaussian curvature of the surface).

Instead of using the functionals $V_{\mu}$ directly, we will mainly
work with the MF densities $v_{\mu}$: 
\begin{equation}
v_{\mu}\equiv V_{\mu}/V_{{\rm Survey}}\;;\;\mu\in\left\{ 0,\ldots,3\right\} \;.\label{eq:MinkDens}
\end{equation}
Here $V_{{\rm Survey}}$ is the volume within the survey boundary.
In the practical calculation of the $v_{\mu}$ discussed in Sec.~\ref{sub:SamplingCode},
$V_{{\rm Survey}}$ is the volume of the region that is more than
two ball radii away from our custom angular mask of Fig. \ref{fig:customMask}.

\subsection{Germ--grain vs. isodensity contours\label{sub:Germ--grain}}

As mentioned, MFs are only defined for extended bodies. For the analysis
of large-scale structure, however, the galaxies are usually treated
as points. For points, all quantities in Eq.~(\ref{eq:MinkRelGeo})
are, a priori, trivial. For a meaningful analysis of large-scale structure,
one therefore has to smooth the point distribution in some manner
to define extended regions. The two most popular choices for defining
extended regions are:
\begin{enumerate}
\item isodensity contours
\item the germ-grain model
\end{enumerate}
The first approach attempts to reconstruct the underlying density
field from the observed galaxies and to define the extended bodies
as the interior of the surfaces of constant density. The value of
this density threshold can then be varied and the functionals are
studied as a function of this parameter.

\begin{figure}
\begin{centering}
\includegraphics[width=0.45\textwidth]{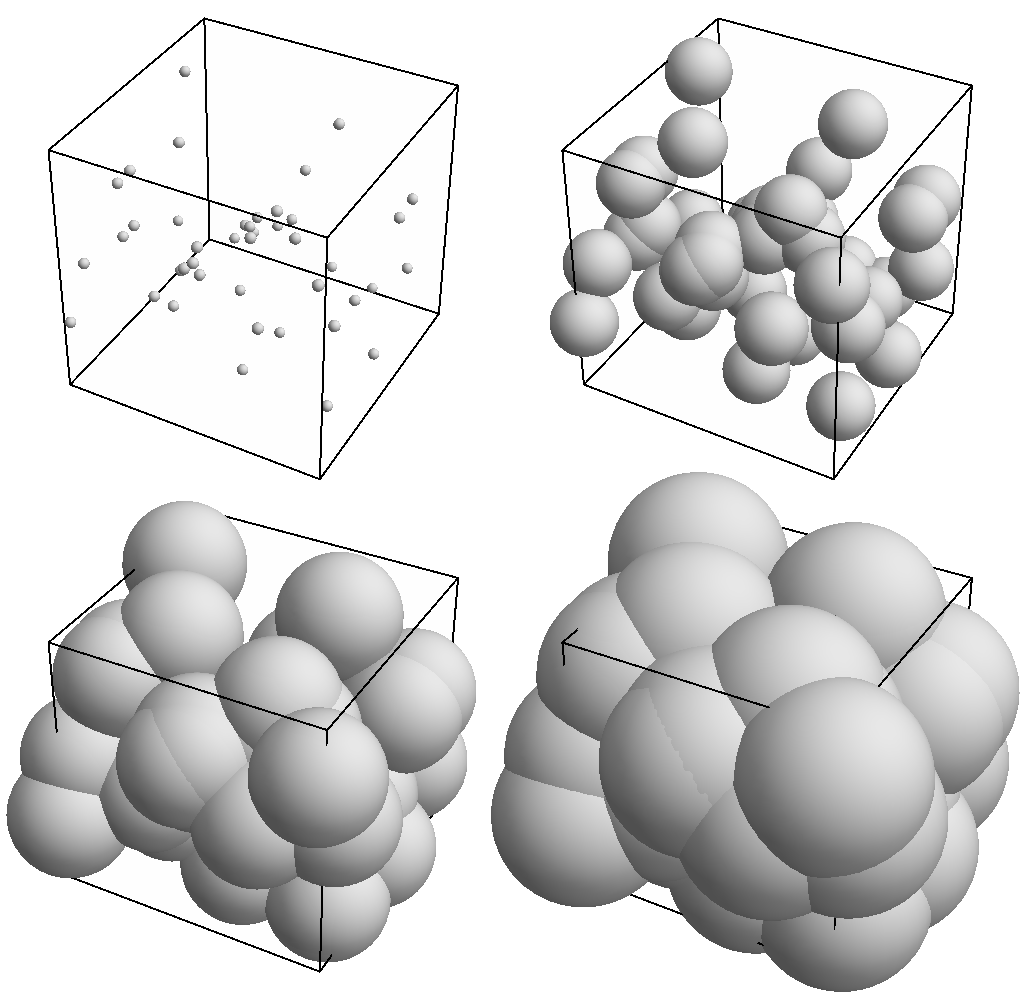}
\end{centering}
\caption{Illustration of the germ-grain procedure to transform a set of galaxies
into an extended body. The galaxy positions $\{\mathbf{x}_{1},\dots,\mathbf{x}_{N}\}$
are surrounded by balls of a common radius $R$. Then the MFs of the
body formed by the union of all the balls are studied as a function
of $R$. The parts of the body outside of the survey region are not
taken into account. With increasing $R$, more and more Balls intersect
and a complex structure develops. Through the single and multiple
intersections of neighboring balls, it encodes information about the
$n$-point correlation hierarchy. Plot from \protect\citet{2014MNRAS.443..241W}.}

\label{fig:boolean} 
\end{figure}

In the second approach, the extended bodies are constructed by surrounding
every galaxy (the germ) by an extended body (the grain), which in
our case is a ball of a certain radius, and calculating the functionals
for the union of all these balls. The diagnostic parameter in this
case is the common radius of the balls. Fig.~\ref{fig:boolean} shows
how a structured body emerges with increasing radius.
 Only those parts of the structure are taken into account that lie
within the survey boundaries, in Fig.~\ref{fig:boolean} illustrated
as a cube.

Even though the basic analysis tool is the same in both cases, the
properties of the two approaches are quite different. The isodensity
contours, at least in theory, are more directly linked the underlying
physical density field, whereas the germ-grain radii are purely diagnostic
parameters that merely specify the scale at which the correlations
are probed. Regarding the temporal evolution, the MFs for isodensity
contours turn out to be relatively independent of the (linear) growth
of the density field, if the value of the threshold density is adjusted
accordingly. On the germ-grain side, there is no corresponding relabeling
and, therefore, as we also shall see below, they depend significantly
on the amplitude of the power spectrum. Besides these conceptual differences,
there are some technical advantages to the calculation of germ-grain
MFs:
\begin{enumerate}
\item The evaluation directly uses the point distribution and does not use
a smoothing procedure that is an art by itself. The use of balls does
not introduce additional (smoothing) parameters.
\item The construction of smooth surfaces can be noisy if the density is
low. The rougher smoothing procedure in the germ-grain model on the
other hand always constructs some well defined surface.
\end{enumerate}
For both smoothing methods there have been detailed analysis how the
values of the measured functionals are related to the parameters of
the underlying density distribution. For the germ-grain case we recall
this relation in the following section.

\subsection{Relation to higher order clustering\label{sub:DensDef}}

Due to the simplicity of the geometrical construction as a union of
intersecting balls, it is relatively straightforward to relate the
germ-grain MFs to correlation properties of the analyzed point distribution.
For the case of a Poisson distribution this has been achieved in \citet{mecke1991euler},
with the only relevant parameter being the density of the random sample.
In the presence of structure, the dependence on higher order correlations
has first been demonstrated in \citet{1994A&A...288..697M}. In the
following, we shall use the form derived in \citet{phdSchmalzing,1999MNRAS.309.1007S}
and detailed in the appendix of \citet{2014MNRAS.443..241W}. In these
works it has been shown that the ensemble average of the germ-grain
MF densities defined in Eq.~(\ref{eq:MinkDens}), for the case of
balls as grains, can be expressed in an analytic manner as a function
of the quantities $\varrho_{0}$ and $\overline{V}_{\mu}$. $\varrho_{0}$
is the ensemble average density of the point distribution studied,
and the $\overline{V}_{\mu}$ are the unnormalized modified MFs from
the DR7 paper \citep{2014MNRAS.443..241W}. The analytic relation reads

\begin{eqnarray}
\av{v_{0}} & = & 1-e^{-\varrho_{0}\overline{V}_{0}}\;,\nonumber \\
\av{v_{1}} & = & \varrho_{0}\overline{V}_{1}e^{-\varrho_{0}\overline{V}_{0}}\;,\nonumber \\
\av{v_{2}} & = & \left(\varrho_{0}\overline{V}_{2}-\frac{3\pi}{8}\varrho_{0}^{2}\overline{V}_{1}^{2}\right)e^{-\varrho_{0}\overline{V}_{0}}\;,\nonumber \\
\av{v_{3}} & = & \left(\varrho_{0}\overline{V}_{3}-\frac{9}{2}\varrho_{0}^{2}\overline{V}_{1}\overline{V}_{2}+\frac{9\pi}{16}\varrho_{0}^{3}\overline{V}_{1}^{3}\right)e^{-\varrho_{0}\overline{V}_{0}}\;,\label{eq:MinkDensDef}
\end{eqnarray}
and provides the implicit definition of the $\overline{V}_{\mu}$.
In the rest of the paper we will drop the explicit ensemble averaging
brackets and use $v_{\mu}\equiv\av{v_{\mu}}$ as a shorthand, as we
never refer to the MF density of one individual realization. We discuss
this aspect in more detail in Sec.~\ref{sub:SamplingCode}.

For a Poisson distribution, the $\overline{V}_{\mu}$ are simply the
MFs $V_{\mu}\left(B\right)$ of the balls of common radius $R$ that
we use to obtain extended bodies: 
\begin{eqnarray}
 &  & V_{0}\left(B\right)=\frac{4\pi}{3}R^{3}\;\;;\;\;V_{1}\left(B\right)=\frac{2}{3}\pi R^{2}\;\;;\nonumber \\
 &  & V_{2}\left(B\right)=\frac{4}{3}R\;\;;\;\;V_{3}\left(B\right)=1\;.\label{eq:MinkBall}
\end{eqnarray}
In the Poisson case Eq.~(\ref{eq:MinkDensDef}) reduces to the result
of \citet{mecke1991euler}. For a point distribution with structure,
the $\overline{V}_{\mu}$ are modified by weighted integrals of the
dimensionless higher-order connected correlation functions $\xi_{n}$.
This modification is what we are interested in in this paper. It is
convenient to define it as a multiplicative dimensionless correction
to the Poisson value, i.e. to study the quantities
\begin{equation}
\eta_{\mu}\equiv\overline{V}_{\mu}/V_{\mu}\left(B\right)\;.\label{eq:etaDef}
\end{equation}
We will refer to $\eta_{\mu}$ as the modified MFs. They can be easily
obtained from the quantities we measure, i.e. the average densities
$v_{\mu}$, by inverting Eqs.~(\ref{eq:MinkDensDef}): 
\begin{eqnarray}
\eta_{0} & = & -\frac{1}{\varrho_{0}V_{0}}\ln\left(1-v_{0}\right)\;,\label{eq:Inverse-Mink}\\
\eta_{1} & = & \frac{1}{\varrho_{0}V_{1}}\frac{v_{1}}{\left(1-v_{0}\right)}\;,\nonumber \\
\eta_{2} & = & \frac{1}{\varrho_{0}V_{2}}\left(\frac{v_{2}}{\left(1-v_{0}\right)}+\frac{3\pi}{8}\frac{v_{1}^{2}}{\left(1-v_{0}\right)^{2}}\right)\;,\nonumber \\
\eta_{3} & = & \frac{1}{\varrho_{0}V_{3}}\left(\frac{v_{3}}{\left(1-v_{0}\right)}+\frac{9}{2}\frac{v_{2}v_{1}}{\left(1-v_{0}\right)^{2}}-\frac{9\pi}{8}\frac{v_{1}^{3}}{\left(1-v_{0}\right)^{3}}\right)\;.\nonumber 
\end{eqnarray}
In the analysis below, we will mainly use these modified MFs instead
of the densities $v_{\mu}$, as they are more directly related to
the correlation properties of the point distribution via (\ref{eq:MinkCorrCon}).
From \citet{phdSchmalzing,1999MNRAS.309.1007S,2014MNRAS.443..241W}
the relation of $\eta_{\mu}$ to the correlations of order $n+1$,
$\xi_{n+1}$, is of the form 
\begin{eqnarray}
\eta_{\mu}\left(R,\varrho_{0}\right) & = & 1+\sum_{n=1}^{\infty}\frac{\left(-\varrho_{0}\right)^{n}}{\left(n+1\right)!}\int\d^{3}x_{1}\dots\d^{3}x_{n}\label{eq:MinkCorrCon}\\
 &  & \!\!\!\!\!\!\!\!\!\!\!\!\!\!\!\! \times\xi_{n+1}\left(0,\bx_{1},\dots,\bx_{n}\right)\frac{V_{\mu}\left(B\cap B_{\bx_{1}}\cap\dots\cap B_{\bx_{n}}\right)}{V_{\mu}\left(B\right)}\;.\nonumber 
\end{eqnarray}
with $V_{\mu}\left(B\cap B_{\bx_{1}}\cap\dots\cap B_{\bx_{n}}\right)$
being simply the MFs of the \emph{intersection} of $n+1$ balls. $B=B\left(R\right)$
denotes a ball around the origin and $B_{\bx}=B_{\bx}\left(R\right)$
a ball around the point $\bx$. In our model all intersecting balls
have the same radius $R$. For general germ-grain models, this is
not a necessary requirement. One could imagine to increase the weight
of some points by assigning them larger radii than other points. Using
elliptic or even more irregular grains would also be possible. However,
this would complicate the interpretation as we expect that it would
lead to a more complicated relation of the functionals to the correlation
hierarchy than the one of Eq.~(\ref{eq:MinkCorrCon}).

The advantage of the $\eta_{\mu}$ of being more directly related
to the correlation properties is counterbalanced by an increase in
the error. In \citet{2014MNRAS.443..241W} we found weaker constraints
when using the $\overline{V}_{\mu}$ as compared to the $v_{\mu}$.
The main reason for this is the additional error that is introduced
by removing the exponential suppression $e^{-\varrho_{0}\overline{V}_{0}}=\left(1-v_{0}\right)$
from $v_{1}-v_{3}$ in (\ref{eq:MinkDensDef}). For $\eta_{1}$, for
example, 
\[
\frac{\delta\eta_{1}}{\eta_{1}}=\frac{\delta v_{1}}{v_{1}}+\frac{\delta v_{0}}{\left(1-v_{0}\right)}\;.
\]
So the relative error for the fraction of the volume not covered by
any ball enters also for the higher functionals. This spoils to some
extent the good precision that can be obtained for $v_{1}-v_{3}$,
because $v_{0}$ has to be measured in the full volume and $v_{1}-v_{3}$
only on the surface. Especially for small $1-v_{0}$ it becomes costly
to determine $v_{0}$ well enough. We will therefore restrict our
analysis of the $\eta_{\mu}$ to parameter regions where $1-v_{0}$
is larger than $\approx0.01$.

\subsection{Density dependence\label{sub:Density-dependence}}

The modified MFs $\eta_{\mu}$ of Eq.~(\ref{eq:MinkCorrCon}) not
only depend on our diagnostic parameter, i.e. the radius $R$ of the
balls that we are using to smooth the galaxy distribution, but also
on the sample density. The rather complicated $R$ dependence is entirely
encoded in the weight functions $V_{\mu}\left(B\cap B_{\bx_{1}}\cap\dots\cap B_{\bx_{n}}\right)/V_{\mu}\left(B\right)$
in the integrals over correlation functions. The dependence on the
mean density of the galaxy sample $\varrho_{0}$, however, is of an
easier form. It is a simple power series in $\varrho_{0}$ 
\begin{equation}
\eta_{\mu}=\sum_{n=0}^{\infty}\frac{c_{\mu,n+1}}{\left(n+1\right)!}\left(-\varrho_{0}V_{0}\left(B\right)\right)^{n}\label{eq:PowerSeriesDecomp}
\end{equation}
with coefficients $c_{\mu,1}=1$ and 
\begin{eqnarray}
c_{\mu,n+1}\left(R\right) & = & V_{0}^{-n}\left(B\right)\int\xi_{n+1}\left(0,\bx_{1},\dots\bx_{n}\right)\times\label{eq:MinkIntegr}\\
 &  & \frac{V_{\mu}\left(B\cap B_{\bx_{1}}\cap\dots\cap B_{\bx_{n}}\right)}{V_{\mu}\left(B\right)}\d^{3}x_{1}\dots\d^{3}x_{n}\nonumber 
\end{eqnarray}
that contain all of the radius and correlation function dependence.
$V_{0}\left(B\right)$ is the volume of a ball $B$ of radius $R$.
We choose to define the second index of $c_{\mu,n+1}$ such that it
directly corresponds to the order of the integrated correlation function. 

The formal split into $\varrho_{0}$ dependence and correlation function
dependence in Eq.~(\ref{eq:PowerSeriesDecomp}) allows us two things:
First, we can remove the Poisson part and the part that depends on
the two-point correlation function, if we measure the sample density
and the two-point correlations independently. This leaves us with
a quantity 
\begin{equation}
\eta_{\mu}^{h}\left(R,\varrho_{0}\right)\equiv\sum_{n=2}^{\infty}\frac{c_{\mu,n+1}\left(R\right)}{\left(n+1\right)!}\left(-\varrho_{0}V_{0}\left(B\right)\right)^{n}\label{eq:DefModMinkHigher}
\end{equation}
that only includes higher-order contributions. This will prove the
significance of their contribution to the MFs in Sec.~\ref{sub:HighSubtract}.
Second, we can extract the integrals over correlation functions by
fitting a polynomial to the density evolution of $\eta_{\mu}\left(\varrho_{0}\right)$.
This allows to define a fitting formula for the form of the MFs in
the SDSS by giving the values and covariances of the coefficients
$c_{\mu,n}\left(R\right)$. We shall use this in Sec.~\ref{sub:SeriesCoefficients}.

We close with the remark that, in general, a finite number of coefficients
is not sufficient to describe the full shape of a power series of
the sort (\ref{eq:PowerSeriesDecomp}). For cooked up point distributions
with small $\xi_{n}$, but a single large $\xi_{m}$ at some higher
$m$, giving only the lower coefficients will not determine $\eta_{\mu}$
well. For the assumed structure of the Universe as the product of
standard structure formation, however, we find that the coefficients
$c_{\mu,n}$ are well behaved and the series $\eta_{\mu}$ converges
reasonably well (see Sec.~\ref{sub:SeriesCoefficients}).

\subsection{Example: Gauss-Poisson clustering\label{sub:Gauss-Poisson}}

\begin{figure}
\begin{centering}
\includegraphics[width=0.45\textwidth]{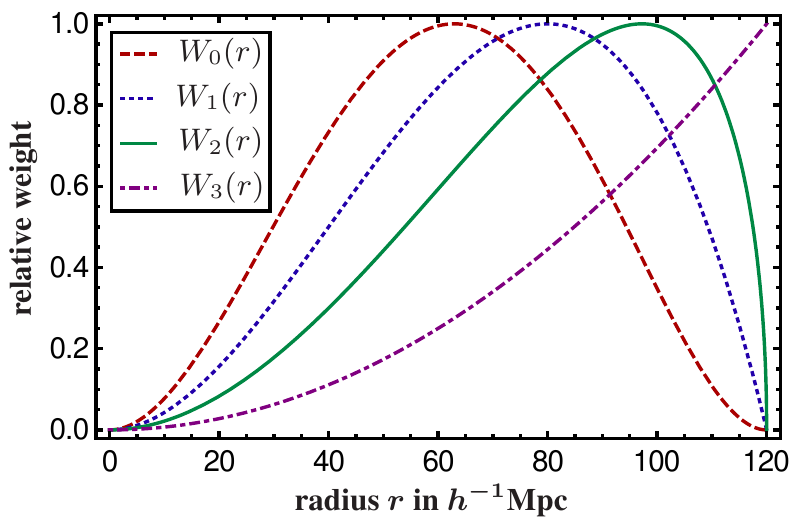}
\end{centering}
\caption{Form of the integration windows $W_{\mu}\left(R,r\right)\equiv V_{\mu}\left(R,r\right)/V_{\mu}\left(R\right)r^{2}$
of Eq.~(\ref{eq:MinkGaussExp-1}), for a ball radius of $R=60\protect\hMpc$.
All functions have been normalized to their maximal value. The functionals
of higher index $\mu$ probe the correlation function at larger distances.
(same plot already in \protect\citet{2014MNRAS.443..241W}) \label{fig:IntegrationWindow}}
\end{figure}

In order to illustrate the relations (\ref{eq:MinkCorrCon}), let
us consider the simplest non-trivial case. In analogy with the continuous
case, one can define a Gaussian point distribution by the absence
of all higher-order connected correlation functions. In this case
the hierarchy of correlations in (\ref{eq:MinkCorrCon}) is truncated
after the linear term and we have

\begin{equation}
\eta_{\mu}=1-\frac{\varrho_{0}}{2}\int\d^{3}x_{1}\xi_{2}\left(0,\bx_{1}\right)\frac{V_{\mu}\left(B\cap B_{\bx_{1}}\right)}{V_{\mu}\left(B\right)}\;.\label{eq:MinkGaussExp}
\end{equation}
The weight functions $V_{\mu}\left(B\cap B_{\bx_{1}}\right)/V_{\mu}\left(B\right)$
are non-zero only for the range of $\bx_{1}$-values for which the
balls intersect. For a coordinate system centered around the origin,
i.e. the center of $B$, the $\bx$-values therefore lie in a sphere
of radius $2R$. A configuration of balls of radius $R$ therefore
probes the correlation function up to a scale $2R$. We shall therefore
plot the radial dependence as a function of the diameter $D=2R$ instead
of the radius. For isotropic correlations (\ref{eq:MinkGaussExp})
becomes 
\begin{equation}
\eta_{\mu}=1-2\pi\varrho_{0}\intop_{0}^{2R}\frac{V_{\mu}\left(R,r\right)}{V_{\mu}\left(B\left(R\right)\right)}\xi_{2}\left(r\right)r^{2}\d r\;.\label{eq:MinkGaussExp-1}
\end{equation}
The shape of the weights $V_{\mu}\left(R,r\right)$ is shown in Fig.~\ref{fig:IntegrationWindow}.\footnote{The equations for the $V_{\mu}\left(R,r\right)$ and their Fourier
transforms can be found in \citet{2014MNRAS.443..241W} (in the Fourier
form of $W_{2}$ in Eq.~(16) there is a typo: the prefactor of the
integral should be $8/3\times R^{3}/k$ instead of $2/3\times R/k$).}
 Let us just recall that the weights are such that $\eta_{0}$ and
$\eta_{3}$ have a particularly interesting form. $\eta_{0}$ directly
contains the amplitude of fluctuations on the scale of the balls $\sigma\left(R\right)$,
\begin{equation}
\eta_{0}=1-\frac{\frac{4\pi}{3}R^{3}\varrho_{0}}{2}\sigma^{2}\left(R\right)\;,\label{eq:V0-sigma}
\end{equation}
and $\eta_{3}$ is a simple integral over the two-point function 
\begin{equation}
\eta_{3}=1-2\pi\varrho_{0}\intop_{0}^{2R}\xi_{2}\left(r\right)r^{2}\d r\;,\label{eq:V3-xi}
\end{equation}
which could therefore be recovered from the derivative of $\eta_{3}$.

Let us pause here and elaborate a bit on the Gauss-Poisson process
as such (see \citet{2001PhRvE..64e6109K} for a more complete discussion
of what follows). It is tempting to think of the truncated quantities
(\ref{eq:MinkGaussExp-1}) as being the MFs of a point distribution
that represents the Gaussian limit of an actual point distribution
that has higher-order correlations. However, in general there will
not be a Gauss-Poisson point process that actually realizes these
MFs, as in order for a Gauss-Poisson process to exist, $\xi_{2}$
and $\varrho_{0}$ have to fulfill certain conditions.

Already in the continuum case, the clustering amplitude is limited
by the fact that the matter density can only be positive. The density
distribution, therefore, is in any case only approximated by a Gaussian.
This approximation of course is, excellent for early times and large
scales, but when matter fluctuations $\sigma\left(R\right)$ grow
too large it breaks down. This is also the point where higher-order
correlations become important.

In the same manner, the clustering amplitude of a Gauss-Poisson point
process is limited. In this case, however, also the mean density of
the point sample $\varrho_{0}$ plays a role. For low enough density
a thinning of the point distribution goes to an approximated Poisson
distribution. This effect can compensate larger fluctuations and it
can be shown \citep{2001PhRvE..64e6109K} that the corresponding constraint
is 
\begin{equation}
\sigma^{2}\left(A\right)\leq\frac{1}{V_{A}\varrho_{0}}=\frac{1}{\overline{N}_{A}}\label{eq:FluctConstraint}
\end{equation}
where $\sigma^{2}\left(A\right)$ are the fluctuations in the number
density of cells $A$ stemming from structure ($\sigma^{2}=\overline{N}_{A}+\overline{N}_{A}^{2}\sigma^{2}\left(A\right)$
are the full fluctuations including the Poisson term), $V_{A}$ is
the volume of one of these cells, and $\overline{N}_{A}$ the mean
number of particles in such a cell. This constraint has two interesting
implications: First, we can already tell from the amplitude of the
two-point function if the field will have a significant amount of
higher order clustering. When we evaluate (\ref{eq:FluctConstraint})
for the structure that we measure in the BOSS galaxies, we find that
the constraint actually is violated. %
{} Second, to study the Gaussian limit on large scales it is not sufficient
to simply increase the domain size (the ball size in our case). In
some cases, where the fluctuations on a region decay faster than its
volume the constraint indicates that the structure for larger cells
becomes more Gaussian. In general, however, one will have to reduce
the number density. In the continuous case, by smoothing over larger
and larger regions, the amplitude of the fluctuations between these
regions decreases and the Gaussian approximation becomes better and
better. In the discrete case, the analogous coarse graining procedure
involves a subsampling of the density field. In the same way that
the number of regions is reduced in the continuous case when averaging
the density over larger cells, individual points of the downsampled
point distribution represent larger regions. We will demonstrate this
effect in Sec.~\ref{sub:MFdensities} where we show the MFs for a
couple of densities. The lower this density becomes, the better the
Gaussian model fits.

In addition to the requirement (\ref{eq:FluctConstraint}), \citet{2001PhRvE..64e6109K}
also showed that the correlation function of a Gauss-Poisson point
process has to satisfy
\begin{equation}
\varrho_{0}\intop_{A}d^{3}x\xi_{2}\left(\left|\bx\right|\right)\leq1\label{eq:xiintconst}
\end{equation}
and cannot be negative

\begin{equation}
\xi_{2}\left(r\right)\geq0\;\;\forall r\in R\;.\label{eq:xiconst}
\end{equation}
Any violation of (\ref{eq:FluctConstraint}) -- (\ref{eq:xiconst})
already implies the existence of higher-order correlations $\xi_{n}$
with $n\geq3$.

We note finally that the condition (\ref{eq:FluctConstraint}) prevents
(\ref{eq:V0-sigma}) and therefore the volume fraction $v_{0}$ in
(\ref{eq:MinkDensDef}) from becoming negative. So we could have derived
a similar constraint on a Gauss-Poisson process from the requirement
that $0\leq v_{0}\leq1$.

\subsection{Practical realization and the Chipmink code\label{sub:SamplingCode}}

The procedure we use to measure the functionals is the same as in
the DR7 analysis \citep{2014MNRAS.443..241W}. As the quantities that
are related to the hierarchy of correlation functions via (\ref{eq:MinkDensDef})
are, strictly speaking, the ensemble averages of the MF densities
$v_{\mu}$, we approximate these averages by creating randomly subsampled
realizations of the point distribution and averaging over their MF
densities. Mathematically speaking there is no guarantee that these
averages should converge to those of an ensemble of independent realizations
(see also \citet{2000ApJ...535L..13K}), but practically the procedure
is working quite well. For studies on small scales one could in principle
use subfields of the full survey as independent realizations and perform
an average that is then more likely to converge (even though it also
requires the assumption of ergodicity and is only valid approximately
as neighboring regions are not completely independent). For the scales
that we are interested in, however, there is only one realization
of the survey and therefore we use the subsampling. For details on
how we generate the samples see the end of Sec.~\ref{sub:Radial}.

The calculation of the $v_{\mu}$ for a given downsampled realization
is performed by \textsc{\footnotesize CHIPMINK} (Code for High--speed
Investigation of Partial Minkowski Functionals), introduced in the
DR7 paper \citep{2014MNRAS.443..241W}. In a first step, it generates
neighbor lists for every galaxy in the sample. Galaxies are considered
as neighbors if they are closer than twice the maximal radius of the
balls. To determine $v_{1}$-$v_{3}$, the code considers galaxies
that are more than $2R$ away from the boundary. Using the list of
neighbors the code calculates the uncovered surface, integrated mean
curvature and Euler characteristic of the ball put around the galaxy.
These partial functionals are then summed to give the global $v_{\mu}$.
The volume used for obtaining the densities from the summed $V_{1}-V_{3}$
is the effective part that is more than $2R$ away from the boundary.

For the volume functional $v_{0}$ another set of neighbor lists is
generated that contains all galaxies within $2R_{max}$ around points
from a random sample filling the same volume as the data, but also
eventual holes in the data that are not treated as internal boundaries.
Around each primary random point we generate $10,000$ random points
in spheres of increasing radius $0<R<R_{max}$. The volume fraction
covered by the balls is then estimated by counting the fraction of
the secondary randoms that have a distance of less than $R$ to any
of the galaxies in the neighbor list of the primary random point.

As compared to the DR7 code, we made a couple of enhancements. Besides
improved memory management, the code can now handle boundaries of
arbitrary shape in ra-dec. It can use data with non-constant density,
uses a larger fraction of points for radii smaller than $R_{max}$,
and samples the $R-\varrho_{0}$-plane more efficiently than the earlier
version. We plan to make also the current version publicly available.

\section{Data reduction}\label{sec:Data}

Having completed the discussion of the methods, we now turn to a description
of the data. The galaxy sample we are analyzing was observed by the
Sloan Digital Sky Survey (SDSS) \citep{2000AJ....120.1579Y} using
the 2.5-meter Sloan Telescope \citep{2006AJ....131.2332G}. Photometric
data used for target selection was obtained using a drift-scanning
mosaic CCD camera \citep{1998AJ....116.3040G} and information on the
camera, photometry and photometric pipeline can be found in \citet{1996AJ....111.1748F,2001ASPC..238..269L,2002AJ....123.2121S,2003AJ....125.1559P,2008ApJ...674.1217P,2010AJ....139.1628D}.
All the photometry was re-processed and released in the eighth data
release \citep{2011ApJS..193...29A}. Based on the so obtained target
catalog, the spectroscopic redshifts are taken by the spectrographs
described in \citet{2013AJ....146...32S}. The final step of spectroscopic
data reduction and redshift determination is presented in \citet{2012AJ....144..144B}.

The data analyzed here is contained in the twelfth data release (DR12)
\citep{2015ApJS..219...12A} of SDSS data and was obtained during the
third stage of SDSS \citep{2011AJ....142...72E} within the Baryon
Oscillation Spectroscopic Survey (BOSS) \citep{2013AJ....145...10D}.
It is described in detail in \citet{2016MNRAS.455.1553R}. The full
data set consists of the spectroscopic redshifts of 836,347 galaxies.
The area probed is divided in a northern and a southern part and two
different target redshift ranges. For our analysis we use the largest
contiguous part of the survey, which is the northern CMASS sample.
Given the technical progress in the boundary treatment of our code,
mentioned in Sec.~\ref{sub:SamplingCode}, it would now also be possible
to use it for the combined sample analyzed in \citet{2016arXiv160703155A}.

The mock catalogs we are using for estimating errors and covariances
were constructed using two different methods. One is based on the
quick particle mesh technique (QPM) described in \citet{2014MNRAS.437.2594W}
and the other one is a hybrid method \citep{2014MNRAS.439L..21K} which
employs what the authors call ``augmented'' second-order Lagrangian
perturbation theory. It was used for generating the MultiDark(MD)-Patchy
mocks \citep{2016MNRAS.456.4156K}. We shall work with both mock catalogs
in our analysis, but concentrate on the MD-Patchy mocks.

\subsection{The SDSS DR12 CMASS data and mock catalogs\label{sub:CMASS}}

\begin{figure}
\begin{centering}
\includegraphics[width=0.45\textwidth]{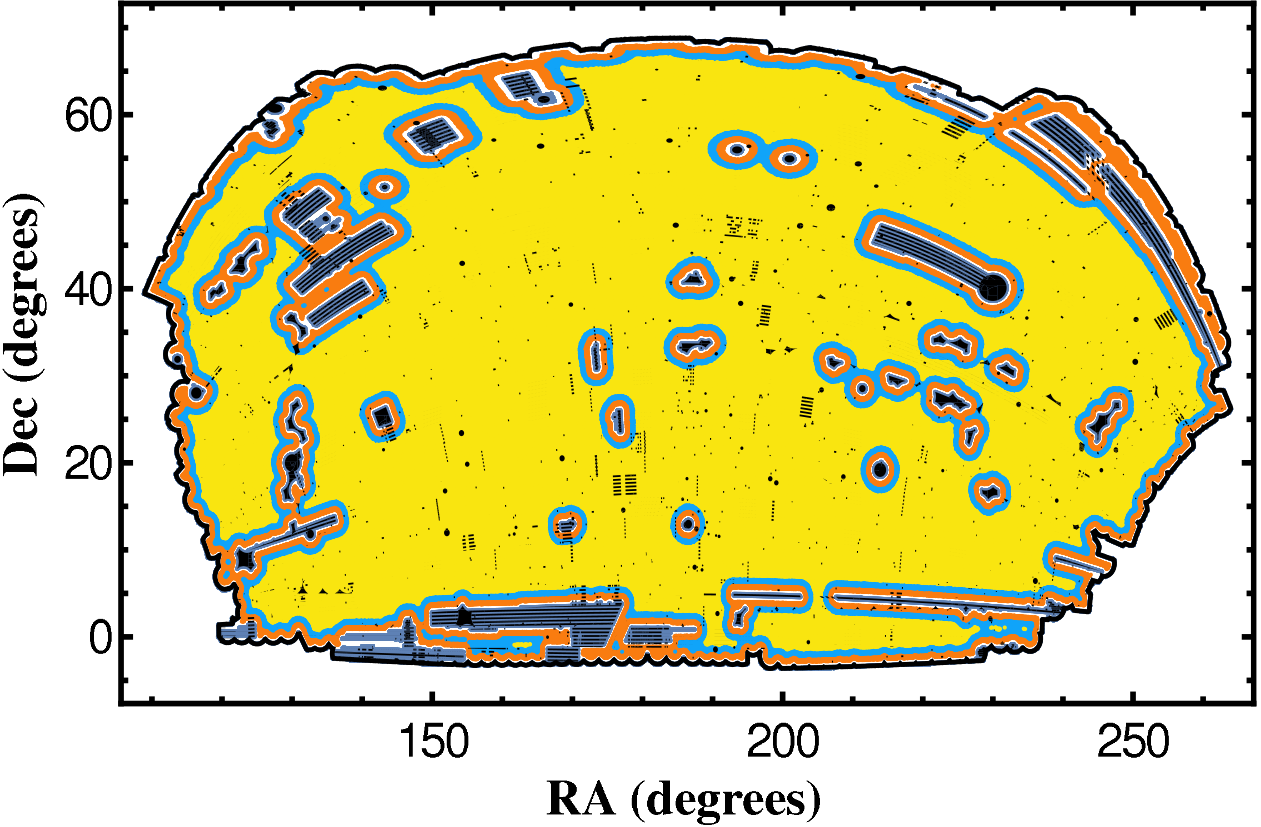}
\end{centering}
\caption{External and internal boundaries. In black are regions excluded by
one of the four veto masks we consider. In blue the subset of the
black regions that we chose to define our boundary. In orange, light
blue and yellow a projection of the regions $18$, $36$ and $54\protect\hMpc$
away from the nearest boundary.\label{fig:customMask}}
\end{figure}
For taking the data of the northern CMASS sample, $685,698$ objects
were targeted over an area of $7,429\,{\rm deg}^{2}$. Out of these,
$607,357$ turned out to be galaxies and obtained redshifts from good
BOSS spectra \citep{2016MNRAS.455.1553R}. They are distributed over
$6,934\,\deg^{2}$. Their redshifts range approximately from $0.4$
to $0.8$. The central region of redshift $0.43$-$0.70$ that is
used in most of the CMASS analysis (see e.g. \citet{2016MNRAS.457.1770C})
contains $568,776$ galaxies.

\begin{figure*}
\begin{centering}
\includegraphics[width=0.9\textwidth]{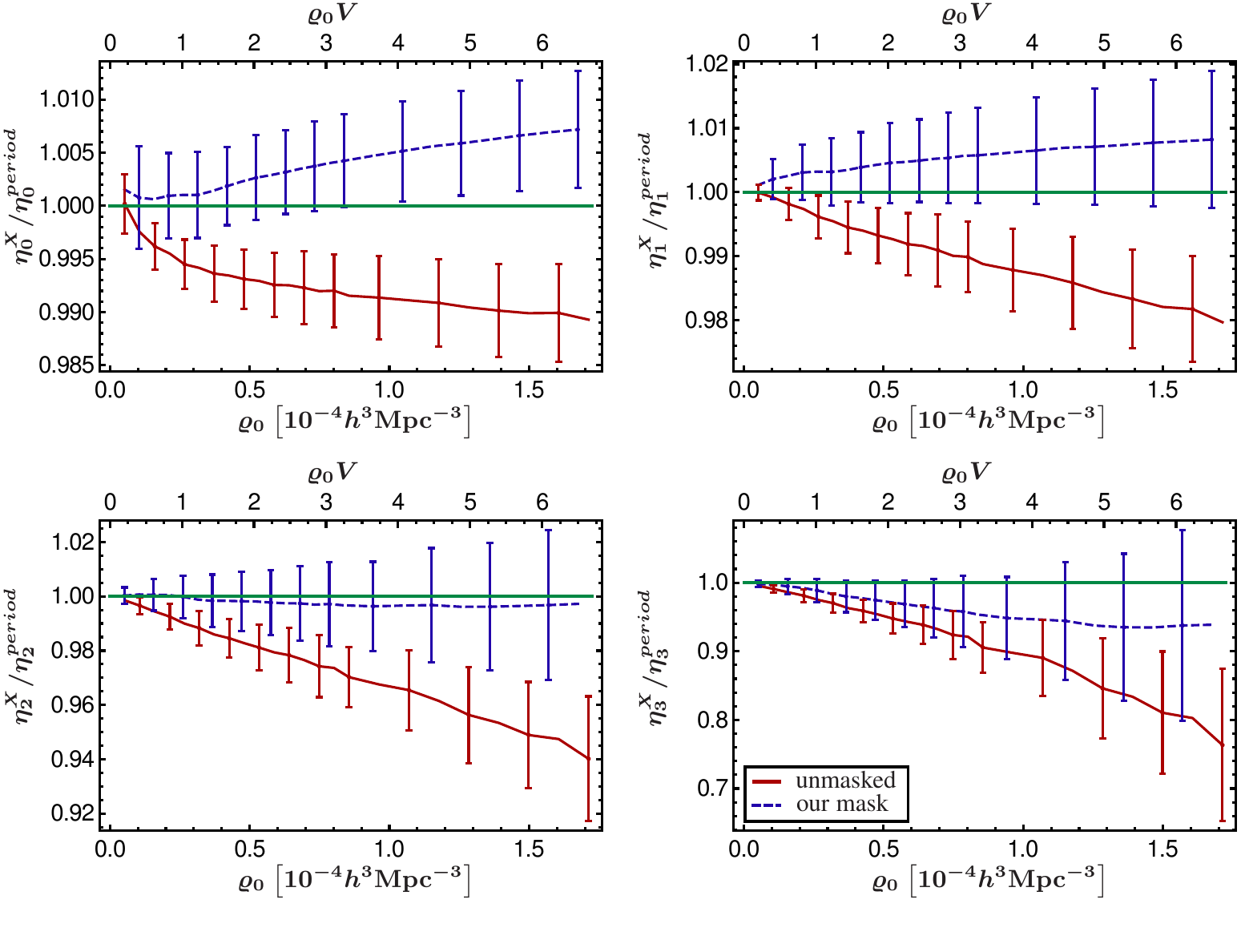}
\end{centering}
\caption{Ratios $\eta_{\mu}^{uncorr}\left(\varrho\right)/\eta_{\mu}^{period}\left(\varrho\right)$
(solid line) $\eta_{\mu}^{corr}\left(\varrho\right)/\eta_{\mu}^{period}\left(\varrho\right)$
(dashed line) for a diameter of $42\protect\hMpc$. $\eta_{\mu}^{uncorr}\left(\varrho\right)$
is determined for the MD-Patchy galaxies after applying the full veto
mask that includes all seven effects discussed in the text. Only the
outer boundaries are taken into account when calculating the $v_{\mu}$.
$\eta_{\mu}^{corr}\left(\varrho\right)$ uses the same galaxies, but
includes also the inner boundaries chosen in Fig.~\ref{fig:customMask}.
$\eta_{\mu}^{period}\left(\varrho\right)$ are derived from cubic
periodic boxes without any boundary. The deviation is largest for
the functionals probing the largest scales $\eta_{3}$ and reduces
to a sub-percent deviation for $\eta_{0}$. For $\eta_{2}$ and $\eta_{3}$
our mask correction is most effective and getting $\eta_{\mu}^{period}\left(\varrho\right)$
within the error bars of $\eta_{\mu}^{corr}\left(\varrho\right)$.
For smaller diameter than $42\protect\hMpc$, the relative deviation
is less important.\label{fig:MaskCorrection}}
\end{figure*}
The MD-Patchy and QPM mock catalogs are constructed such that they
trace the number density of the data. They are available for the central
range of $0.43$-$0.70$. Their angular distribution fills the full
survey geometry. For matching the angular distribution of the data
as well, regions of bad data quality or missing observations can be
masked out. The QPM mocks contain on average $651,316$ galaxies.
After application of the full veto mask the average number is $576,517$.
For the MD-Patchy mocks, the unmasked number of galaxies is $674,880$
and the masked one $591,192$.

Our custom radial selection of a redshift range of $z=0.450-0.595$,
explained in Section \ref{sub:Radial}, reduces the number of galaxies
that we use to $410,615$.

\subsection{Correcting for the survey geometry\label{sub:Ceometry}}

In order to compare the data to the galaxy samples created in the
mock catalogs, it would be sufficient to calculate the MFs for the
masked mocks and the data. However, ideally we would like to derive
quantities that are independent of effects of the survey boundary.
This is not necessary for the direct comparison, but desirable for
later comparison of the data with other simulations without having
to apply the survey mask to them. When estimating the correlation
function, the effect of boundaries (inner boundaries such as holes
as well as the outer boundary) can be taken into account by using
random samples with the exact same boundaries as the data. The excess
of data over random pairs in a given bin is then solely due to structure.

In the case of the MFs, unfortunately there is no simple analogue
to this approach. We tried to correct the MFs of the masked mocks
with the MFs of a random distribution and the same mask. A comparison
with the unmasked mocks showed that this correction is not effective
in the MF case. In order to minimize the influence of boundaries,
we therefore choose to restrict our analysis to regions with good
data quality. This is achieved by using only galaxies that are more
than $2R$ away from the closest boundary, where $R$ is the common
radius of the balls around every galaxy (see Sec.~\ref{sub:Germ--grain}).
Of course, especially for large $R$, this reduces the volume probed
and so the number of galaxies that can be used. We therefore have
to find a compromise as to which boundaries we include and which we
can ignore without affecting the result too much.

As discussed in detail in \citet{2016MNRAS.455.1553R}, there are at
least seven different effects that lead to bad or missing data. For
each effect there is a veto mask that can be used in the correlation
function analysis. The effects are:
\begin{itemize}
\item The central bolt obscures the center of the plate
\item Ly--$\alpha$ targets are given priority in fiber assignment
\item No observations around bright stars
\item No observations around bright objects
\item Non-photometric conditions
\item Spectra with bad seeing
\item Spectra degraded by extinction
\end{itemize}
The regions obscured by the first three effects are quite small individually,
but relatively evenly distributed over the sky. Including them, and
especially the $2R$ region around them, would reduce the usable area
by a large amount, so we can not correct for their effect. The fourth
effect is also present all over the sky, however, the obscured regions
have a wide range of sizes. Therefore we can exclude the largest ones.

\begin{figure}
\begin{centering}
\includegraphics[width=0.45\textwidth]{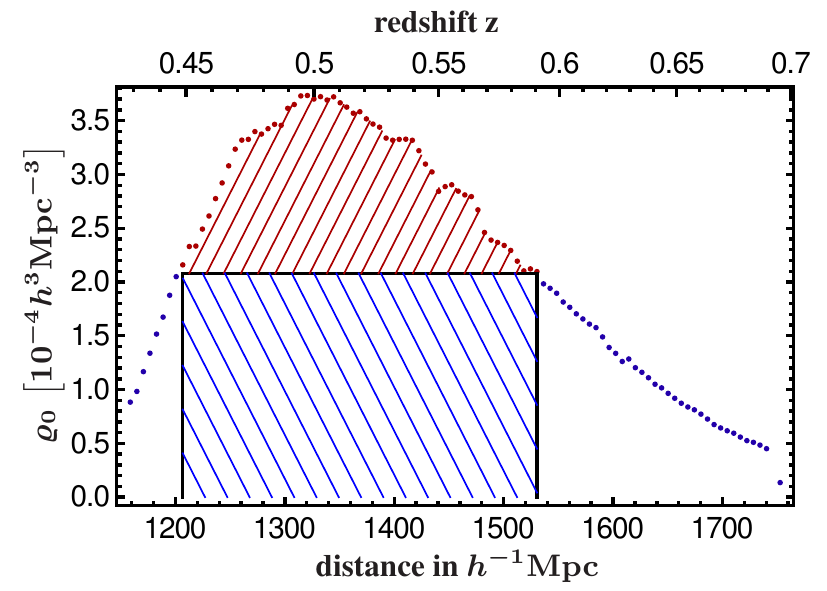}
\end{centering}
\caption{Radial selection: The full redshift range analyzed is $z=0.450-0.595$.
The points indicate the minimal value of the density of the 1000 QPM
mocks in the respective bin. The reference density with respect to
which we define our downsampling is $2.08\times10^{-4}\protect\hDens$.
This combination of reference density and redshift range maximizes
the number of downsampled galaxies in the boxed blue shaded region.
We also use the points above the reference density by downsampling
always from the full density.\label{fig:RadialSelection}}
\end{figure}

The combination of the lower four masks is shown as the black regions
in Fig.~\ref{fig:customMask}.
 Of those regions we select those with an area on the sky of more
than about $0.8\,{\rm deg}^{2}$. We densely sample these regions
and their boundaries with points that are shown in blue. These are
used to determine the distance of a galaxy from the nearest boundary.
The projected holes created by ball diameters $2R$ of $18$, $36$
and $54\,\hMpc$ are shown as the regions not filled by orange, blue
and yellow color. At these diameters the effective volume left is
$75\%$, $57\%$ and $42\%$, respectively. For the largest scales
probed of $120\hMpc$, only $8\%$ remains. %

As the mask we constructed does not take into account all imperfections,
we have to test the extent to which the result still depends on the
survey boundaries. In order to study this, we removed the galaxies
from the mocks that fall into one of the seven masks described above.
Then we determined the distance of the remaining galaxies from our
boundaries that are shown in Fig.~\ref{fig:customMask}. We calculated
their germ-grain MFs and those of the galaxy in the corresponding
cubic mocks periodically wrapped to remove any boundary. The resulting
comparison is shown in Fig.~\ref{fig:MaskCorrection}. 
It also contains the case where we use the full veto mask to remove
the unobserved galaxies, but only take the outer boundaries into account
in the calculation of the MFs. We find that our custom mask greatly
improves the match with the boundaryless case as compared to not taking
inner boundaries into account. After the correction the periodic MFs
are now within the error bars of the sky MFs.

The sky values of $\eta_{0}$ in Fig.~\ref{fig:MaskCorrection} have
been corrected for a systematic $0.5\%$ offset from the periodic
case. We shall apply the same correction throughout the rest of the
paper. As the offset has been derived from a comparison of the sky
mocks to the periodic mocks, one could think of using the deviations
of Fig.~\ref{fig:MaskCorrection} to apply density and scale dependent
corrections to both the sky mocks and the data, to empirically remove
the effect of the boundary. We decided not to apply this correction
as it would make the results for the MFs of the data completely dependent
on this particular set of mocks.

From Fig.~\ref{fig:MaskCorrection}, we also see that our correction
is largest for the fourth functional $\eta_{3}$ and least important
for $\eta_{0}$. This reflects the property of the integration kernel
of Fig.~\ref{fig:IntegrationWindow}. With increasing index $\mu$,
the functionals probe larger and larger scales and are hence more
sensitive to the large scale boundary effects we removed.

\subsection{Radial selection and density binning\label{sub:Radial}}

In addition to the angular selection, we also apply some radial cuts.
As described in Sec.~\ref{sub:Density-dependence} we want to study
the MFs of the galaxy distribution as a function of ball radius, but
also sample density. A controlled variation of sample density requires
a roughly constant density throughout the sample. Otherwise the summation
of partial MFs (see Sec.~\ref{sub:SamplingCode}) from different
regions of the survey would combine $v_{\mu}$ values effectively
measured at different densities, removing the clear distinction between
the $\varrho_{0}$ and the $\xi_{n}$ dependence in Eq.~(\ref{eq:MinkDensDef}).

To generate a homogeneous sample we downsample the density to a common
value for all radial shells. The concept is shown in Fig.~\ref{fig:RadialSelection}.
\begin{figure}
\begin{centering}
\includegraphics[width=0.45\textwidth]{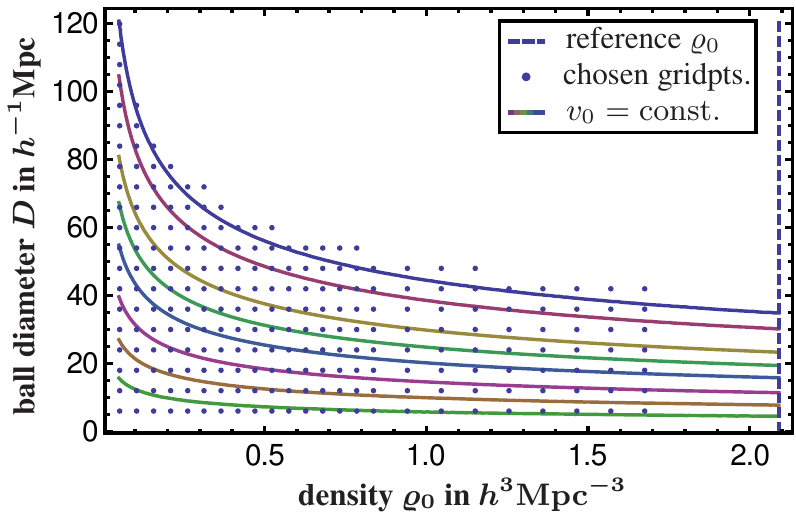}
\end{centering}
\caption{Choice of points in the $D$-$\varrho_{0}$-space. For the smaller
diameters there are $24$ density points, for the largest diameter
used in the analysis $66\protect\hMpc$, there are $7$. The lines
of constant $v_{0}$ are calculated for a Poisson point process and
correspond to \{0.99, 0.95, 0.75, 0.55, 0.35, 0.15, 0.05, 0.01\} from
top to bottom.\label{fig:DensityGrid}}
\end{figure}
 There, we plot the binned density distribution of the QPM mocks.
For each bin we show the minimum of the 1000 QPM mocks. Using this
density distribution, we choose a reference density such that the
number of points in the constant density region of the considered
redshift range is maximal. This density is $2.08\times10^{-4}\hDens$
and corresponds to $296,738$ galaxies. The total number of galaxies
in this redshift range is $410,615$. As we used the minimum of all
mocks, we can be sure that we can downsample from this reference density
and obtain the same constant density for every single mock realization.

The sampling points in radius and density that we used are shown in
Fig.~\ref{fig:DensityGrid}. For ball diameters up to $42\hMpc$,
there are $24$ density points ranging from $2.5\%$ to $80\%$ of
the reference density. For diameters above $42\hMpc$ we decrease
the maximal density and therefore the number of density points. The
reason for this is two-fold: First, as discussed in Sec.~\ref{sub:DensDef},
for very small fractions of uncovered volume (i.e. large radius and/or
high density), the precision of the determination of $v_{0}$ is no
longer high enough to give reasonable error bars for the derived quantities
$\eta_{\mu}$. Second, the runtime is much worse if the galaxies have
a large numbers of neighbors. The density cutoff in Fig.~\ref{fig:DensityGrid}
is therefore roughly chosen along the lines of constant $v_{0}$.
Fig.~\ref{fig:DensityGrid} includes these lines for the Poisson
case, which we used for a first estimate of reasonable density ranges.

From the range of diameters shown in Fig.~\ref{fig:DensityGrid},
we choose $18\hMpc$ to $66\hMpc$ for our analysis. For smaller scales
than $18\hMpc$ we expect to pick up too much effects from the ignored
holes in the survey and for larger scales than $66\hMpc$, the volume
far enough from the boundary becomes too small.

\begin{figure}
\begin{centering}
\includegraphics[width=0.45\textwidth]{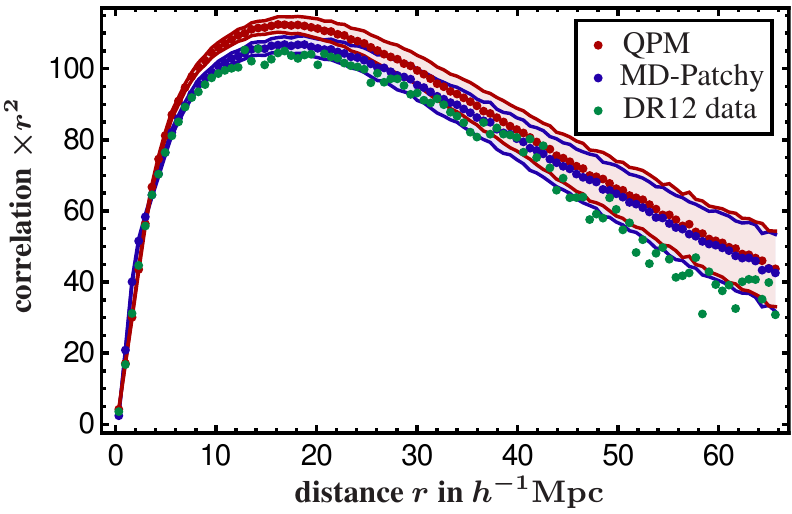}
\end{centering}
\caption{Average correlation functions for the QPM and MD-Patchy mocks compared
to the correlation function of the DR12 data. The shaded region shows
the $1\sigma$ scatter of the mocks.\label{fig:MeanCorrFun}}
\end{figure}

For our independent measurement of the correlation functions of the
data and the mocks, we chose to use those regions that are more than
$36\hMpc$ away from the closest boundary that we treat. The structure
that we measured for these regions is shown in Fig.~\ref{fig:MeanCorrFun}.

The concrete Monte Carlo procedure we use is then to randomly select
galaxies from the full number of galaxies in our redshift range (and
\textit{not} a pre-downsampled reference density as in the DR7 paper)
for each $D-\varrho_{0}$-point of the grid in Fig.~\ref{fig:DensityGrid},
to obtain a sample with the constant target density in each radial
bin. We then calculate the MFs for this configuration. We repeat the
random selection and average the resulting MFs over these different
realizations of the subsampled point distribution. The number of realizations
is largest for the smallest densities and ranges from $617$ for a
fraction of $2.5\%$ to $10$ for a fraction of $80\%$.

\subsection{Weights\label{sub:Weights}}

In addition to the regions that could not be observed or had bad data
quality and that therefore had to be masked out, there are also imperfections
in the rest of the observed survey volume. Even though the completeness
of the survey is high, see \citet{2016MNRAS.455.1553R}, there are
still galaxies that have not obtained robust redshifts (no-z). In
addition due to the physical size of the fibers of the spectrograph,
galaxies separated by less than $62^{\prime\prime}$ can not be observed
simultaneously (close pairs). In the correlation function analysis,
these holes are accounted for by upweighting nearby galaxies. In our
case this can not be done as easily. In principle one could use imaging
data for the angular position and the redshift of nearby galaxies
to mimic the procedure in the correlation function case and fill in
the holes. However, as we are already ignoring a large number of holes
and looking at relatively large scales, we decided not to take these
holes into account and to mask them in the mocks consistently.

However, we include another class of weights correcting for spurious
density fluctuations due to stellar density and seeing. For the correlation
function analysis, these have been combined into a systematic weight
(see \citet{2016MNRAS.455.1553R}). We use this systematic weight to
adjust the probability of selecting a galaxy when we downsample to
a given target density. As the modifications are below $10\%$ and
not factors of $2$ to $7$ as in the case of the no-z and close pairs
weights, the fluctuations can be accounted for relatively easily by
this procedure.

\section{Non-linear clustering as seen by the Minkowski Functionals}\label{sec:HigherOrders}

In this section we first present the measured values for the observables
$v_{\mu}$ introduced in Sec.~\ref{sub:DensDef}. We study their
radial and density dependence based on the MD-Patchy mocks. Then we
turn to the analysis of the higher-order contributions. For this analysis,
we need the $\eta_{\mu}$ that are obtained from the measured $v_{\mu}$
through Eqs.~(\ref{eq:Inverse-Mink}). From the $\eta_{\mu}$ we
can then simply subtract the two-point part of the expansion (\ref{eq:MinkCorrCon})
to extract the $\eta_{\mu}^{h}$ part (\ref{eq:DefModMinkHigher})
that only depends on higher orders. We will see that it contributes
significantly to the full $\eta_{\mu}$. Finally we analyze the coefficients
$c_{\mu,n}$ of (\ref{eq:MinkCorrCon}) order by order and compare
them to correlation function integrals.

\subsection{The functional densities measured\label{sub:MFdensities}}

\begin{figure*}
\begin{centering}
\includegraphics[width=0.9\textwidth]{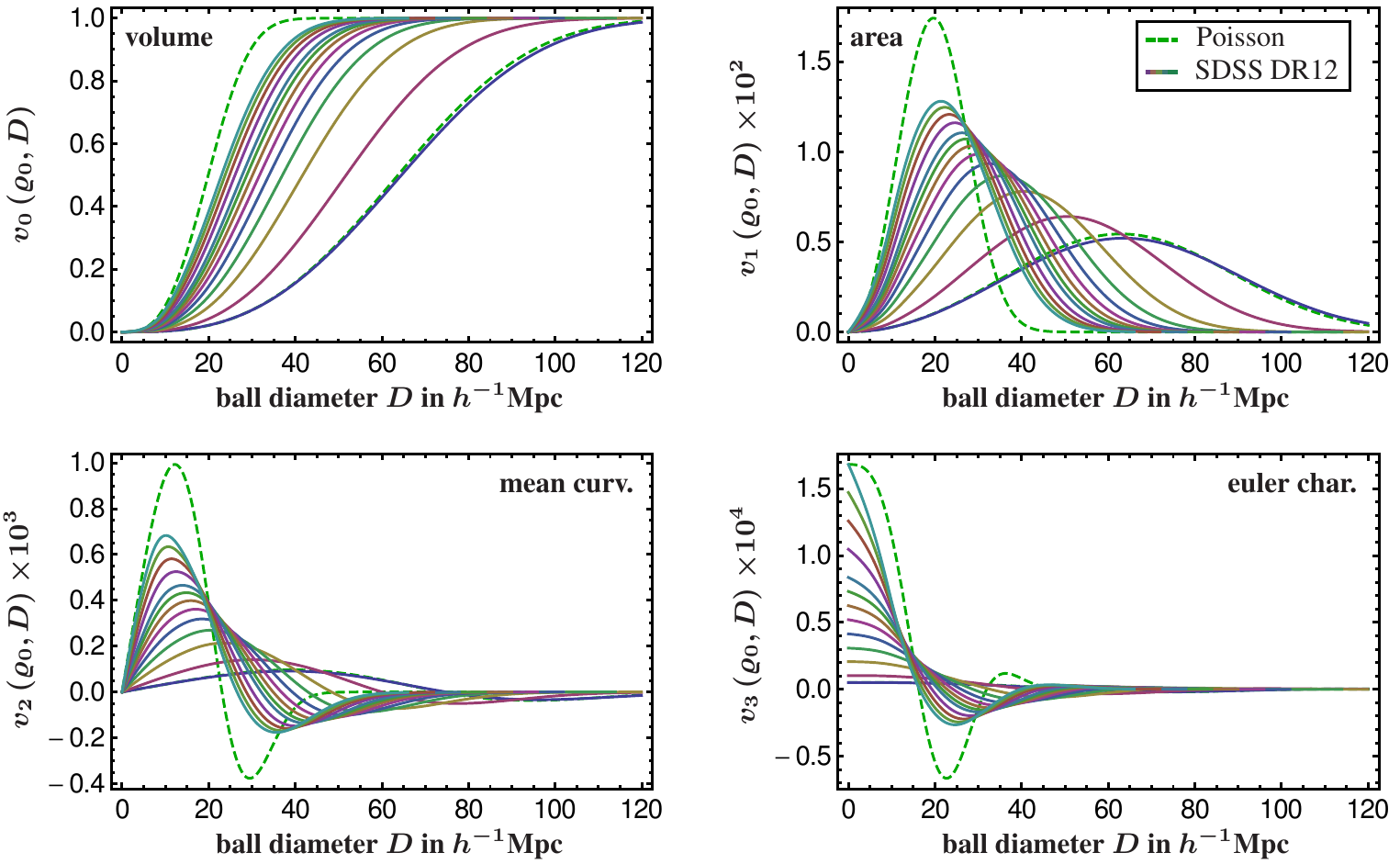}
\end{centering}

\caption{MF densities $v_{\mu}$ from Eq.~(\ref{eq:MinkDensDef}) of the SDSS
DR12 MD-Patchy mocks as a function of the diameter of the balls for
$13$ sample densities ranging from $5.2\times10^{-6}\protect\hDens$
to $1.7\times10^{-4}\protect\hDens$. Higher density curves lie above
lower density ones at small diameters. For the two most extreme densities
the plots show as dashed lines the corresponding curves for a Poisson
point distribution. \label{fig:MinFunDens}}
\end{figure*}

As mentioned in Sec.~\ref{sub:SamplingCode}, the quantities calculated
by \textsc{\footnotesize CHIPMINK} are the the MF densities $v_{\mu}$.
Their dependence on the ball diameter is shown in Fig.~\ref{fig:MinFunDens}
and based on the mean of $30$ MD-Patchy realizations. We only used
$30$ mocks in this case because Fig.~\ref{fig:MinFunDens} is mainly
for illustration and the large range of diameters is not optimal for
the evaluation of the $\eta_{\mu}$ as discussed at the end of Sec.
\ref{sub:DensDef}.
In order to illustrate also the density dependence, we plot $12$
different densities ranging from $2.5\%$ to $80\%$ of our reference
density of $2.08\times10^{-4}\hDens$ (the same density points as
in Fig.~\ref{fig:DensityGrid}, however with an increased maximal
radius). For the most extreme cases of $2.5\%$ and $80\%$, we also
plot the MFs of a Poisson point process as dashed lines. For the high
density case, the functionals for the mocks are clearly non-Poissonian,
whereas in the low density case the downsampling erased already a
lot of the structure and the difference to a Poisson sample is less
important.

Let us qualitatively describe the difference for the Euler characteristic
$v_{3}$ ($v_{3}=\chi=K-H+C$). The faster drop of the functional
with structure indicates that the individual balls aggregate earlier,
which reduces the number of independent components $K$ faster than
for the Poisson case. Then the holes $H$ in the union of multiple
balls prevail up to larger radii and in the end there are less cavities
$C$ but at larger radii.

\begin{figure*}
\begin{centering}
\includegraphics[width=0.9\textwidth]{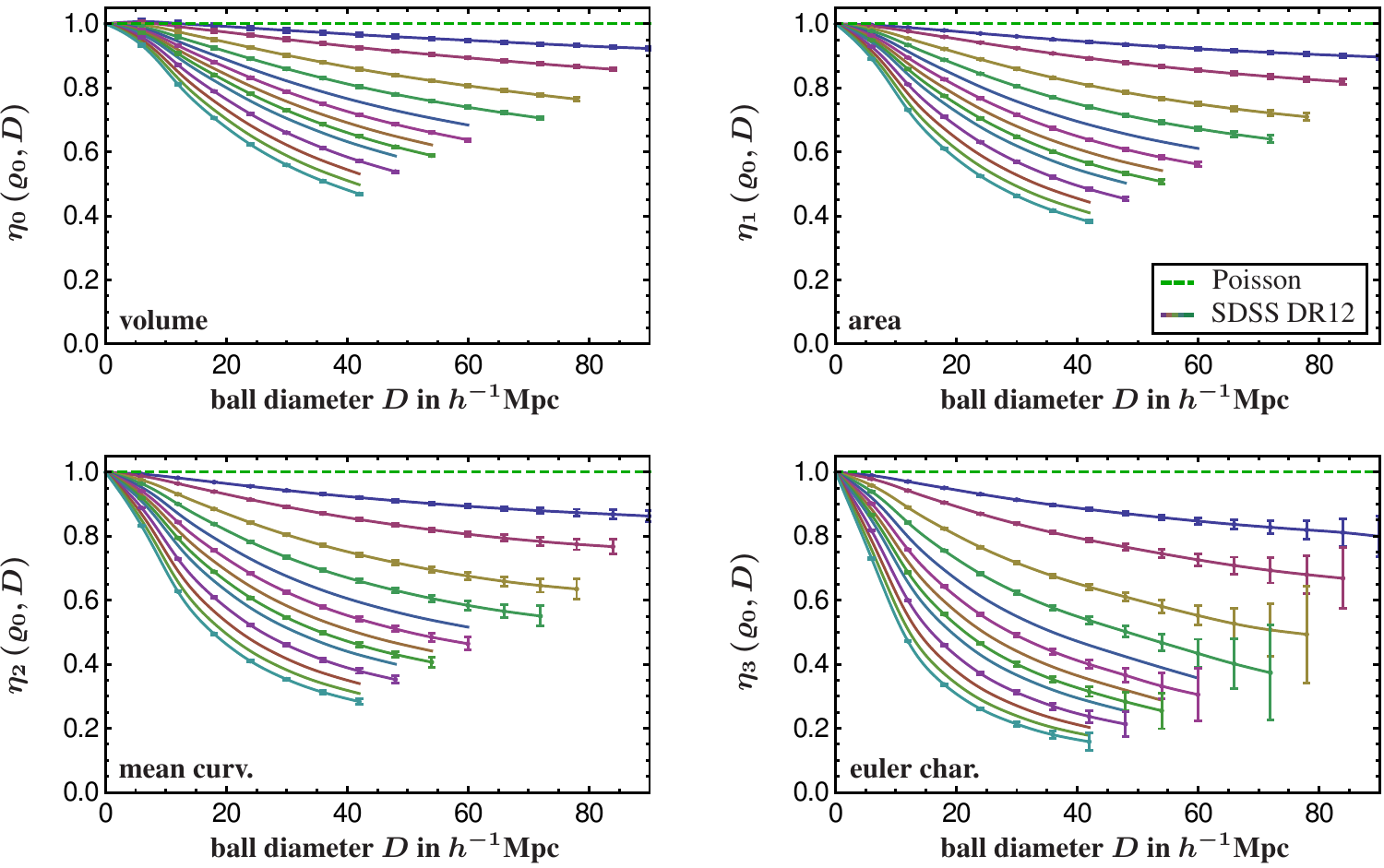}
\end{centering}

\caption{Modified MFs $\eta_{\mu}$ from Eq.~(\ref{eq:MinkCorrCon}) evaluated
for $900$ SDSS DR12 MD-Patchy mocks and plotted as a function of
the diameter of the balls. The lines are the direct transformation
of the $v_{\mu}$-lines of Fig.~\ref{fig:MinFunDens} by Eq.~(\ref{eq:Inverse-Mink}).
The $13$ lines correspond to the same densities as in Fig.~\ref{fig:MinFunDens}.
High densities lie below low densities. The dashed Poisson line is
equal to $1$, thanks to the normalization of the $\eta_{\mu}$. The
$\eta_{\mu}$ are the quantities that are more directly related to
the structure of the sample via the integrated correlation functions
of Eq.~(\ref{eq:MinkCorrCon}) and that we will therefore use in
the following. \label{fig:ModMinkFun}}
\end{figure*}

For our quantitative analysis in Sec.~\ref{sec:HigherOrders} we
will mainly use the modified MFs $\eta_{\mu}$, Eq.~(\ref{eq:MinkCorrCon}).
Therefore, we use Eqs.~(\ref{eq:Inverse-Mink}) to transform from
the initial observables $v_{\mu}\left(\varrho_{0},R\right)$ to $\eta_{\mu}\left(\varrho_{0},R\right)$.
Fig.~\ref{fig:ModMinkFun} shows the dependence of the $\eta_{\mu}$
on the ball diameter for the same densities as in Fig.~\ref{fig:MinFunDens},
but now for 900 MD-Patchy mocks.
 In these plots the convergence of the Functionals to those of the
Poisson point process with decreasing sample density is more visible
than in Fig.~\ref{fig:MinFunDens}. However, given the small error
bars, even for the lowest density of $5.2\times10^{-6}\hDens$ the
structure is well detected.

The differing number of radial points for different densities illustrates
our grid of points sampling the $D-\varrho_{0}$ plane of Fig.~\ref{fig:DensityGrid}.
For the highest density, $7$ radial bins of $\delta R=3\hMpc$ are
possible until the survey volume is significantly filled, for the
lowest density there are $20$ bins (of which only the first $15$
are shown here).

\begin{figure*}
\begin{centering}
\includegraphics[width=0.9\textwidth]{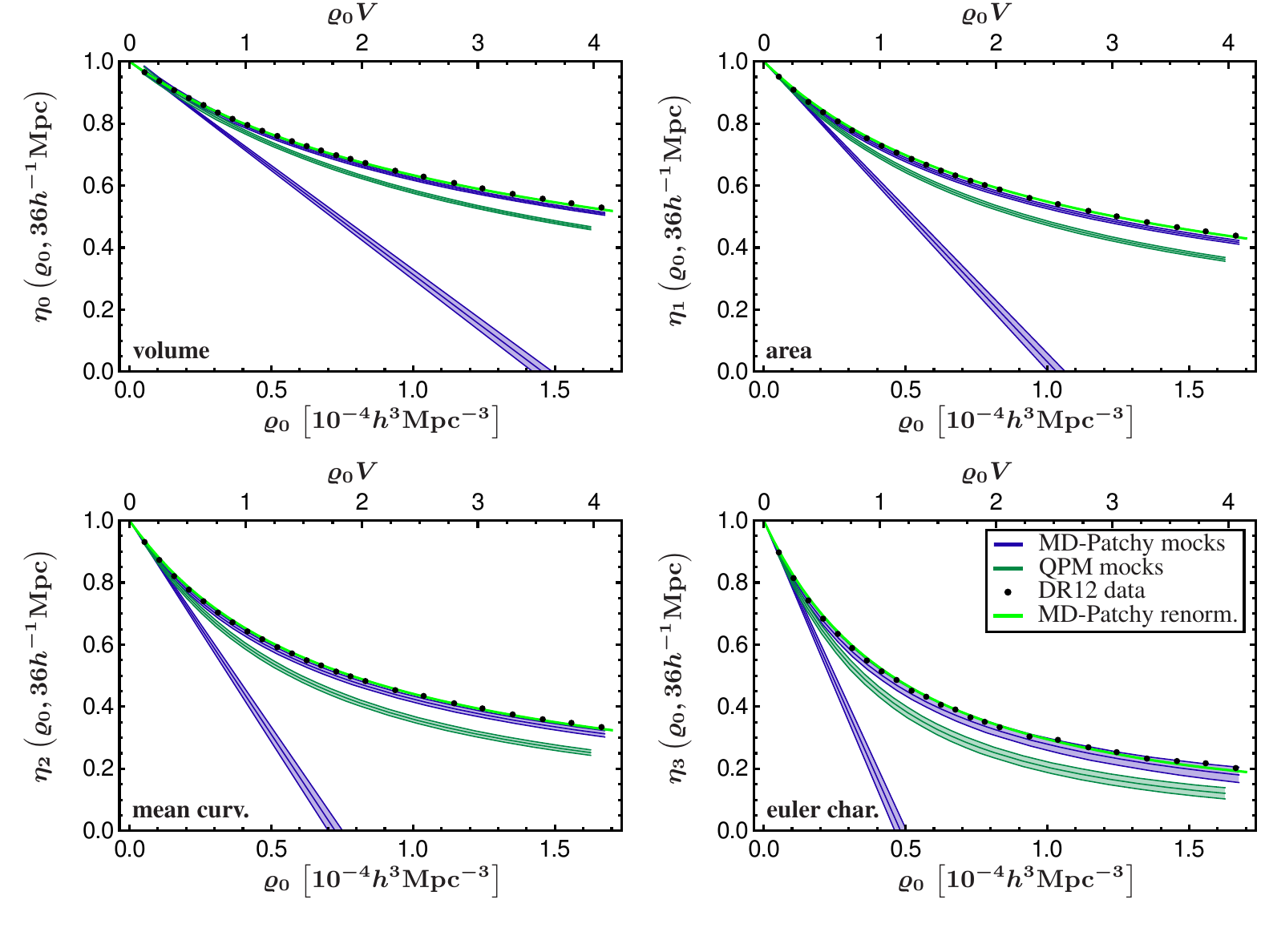}
\end{centering}

\caption{The modified MFs $\eta_{\mu}$ of Fig.~\ref{fig:ModMinkFun}, but
now as a function of the sampling density for a single diameter of
$36\protect\hMpc$. The $\varrho_{0}$ parameter direction has the
advantage to be simply described by a power series in $\varrho_{0}$,
see Eq.~(\ref{eq:PowerSeriesDecomp}). The curved bands give $1-$
and $2-\sigma$ contours of the $800$ QPM (lower, blue band) and
$900$ MD-Patchy (upper, red band) mocks. For comparison the integrated
two-point components derived independently from the same mocks are
given by the straight $1-\sigma$ bands. The green line shows the
expected change of the average MD-Patchy line for a $5\%$ decrease
of $\sigma_{8}^{2}$. \label{fig:ModMinkDensDep}}
\end{figure*}

\begin{table}
\begin{tabular}{ccccccccc}   \hline\hline \vspace{-0.3cm} \\ & \multicolumn{2}{c}{QPM $\eta_{\mu}(\varrho_{0})$} \hspace{-0.3cm} & &  \multicolumn{2}{c}{MD-P. $\eta_{\mu}(\varrho_{0})$} \hspace{-0.3cm} & &  \multicolumn{2}{c}{MD-P. $\eta^{h}_{\mu}(\varrho_{0})$} \\\vspace{-0.3cm} \\ & $\chi^2$ & $\sigma_G$ & & $\chi^2$ & $\sigma_{G}$ & & $\chi^2$ & $\sigma_{G}$ \\  \noalign{\smallskip}\hline\noalign{\smallskip}  $\eta_0$ & 1810 & 41.1 & & 171 & 10.1 & & 29.8 & 1.31 \\  $\eta_1$ & 1120 & 31.7 & & 88.1 & 5.93 & & 32.1 & 1.54 \\  $\eta_2$ & 454 & 19.1 & & 46.9 & 2.93 & & 29.6 & 1.28\\  $\eta_3$ & 182 & 10.6 & & 29.7 & 1.30 & & 23.5 & 0.694 \\ \noalign{\smallskip}\hline \end{tabular}

\caption{Example of $\chi^{2}$ values for the deviation of the data from the
mocks for a single ball diameter of $36\protect\hMpc$. The left table
is derived from the full density evolution of $\eta_{\mu}$ for the
QPM mocks (green band compared to black data points in Fig.~\ref{fig:ModMinkDensDep}).
The central table is the same, but for the MD-Patchy mocks (blue band
in Fig.~\ref{fig:ModMinkDensDep}). The right table quantifies the
deviation of the $\eta_{\mu}^{h}$ of the data from the MD-Patchy
mocks in Fig.~\ref{fig:ModMinkSubtract}. For all tables, $\chi^{2}$
has 24 degrees of freedom. The tables confirm the results of the figures
that the MD-Patchy mocks are describing the data much better than
the QPM mocks and that the higher-order MD-Patchy and data part agree
quite well at this radius.\label{tab:ChiSqMockDat}}
\end{table}

\subsection{Higher-order correlations in the SDSS galaxy data\label{sub:HighSubtract}}

We will now turn to the comparison of the density dependence of the
modified MFs $\eta_{\mu}$ in the mocks and in the SDSS data. Instead
of probing the grid of Fig.~\ref{fig:DensityGrid} for a fixed density
as in Fig.~\ref{fig:ModMinkFun}, we show in Fig.~\ref{fig:ModMinkDensDep}
the $\eta_{\mu}$ as a function of the sampling density for a ball
diameter of $36\hMpc$.
The form of $\eta_{\mu}$ in this direction of parameter space is
easier to describe than in the $D$ direction, due to the power series
form of the $\varrho_{0}$-dependence, see Eq.~(\ref{eq:PowerSeriesDecomp}).
In the plots of Fig.~\ref{fig:ModMinkDensDep}, the Poisson-case
corresponds to the horizontal line at $1$, thanks to the normalization
of the $\eta_{\mu}$ in Eq.~(\ref{eq:MinkCorrCon}). The data are
clearly different from $1$, showing that there is structure in the
galaxy distribution. 

The straight lines that depart from the Poisson case at zero density
show the $\eta_{\mu}$ truncated after the term linear in $\varrho_{0}$,
see Eq.~(\ref{eq:MinkGaussExp}). They have been calculated by measuring
the two-point correlation functions for each of the $800$ QPM and
$900$ MD-Patchy mocks and integrating them with the corresponding
weight functions $V_{\mu}\left(R\right)$ of Eq.~(\ref{eq:MinkGaussExp-1}).
The plot shows the average and the $1-\sigma$ fluctuations between
the different mock realizations. The average of the correlation functions
that formed the basis of these lines is shown in Fig.~\ref{fig:MeanCorrFun}.
The clear deviation of the full $\eta_{\mu}$ of the data and the
mocks from these lines shows that the integrals over higher-order
correlation functions in the series expansion of $\eta_{\mu}$ actually
do play a significant role for the structure in the SDSS galaxies.

\begin{figure*}
\begin{centering}
\includegraphics[width=0.9\textwidth]{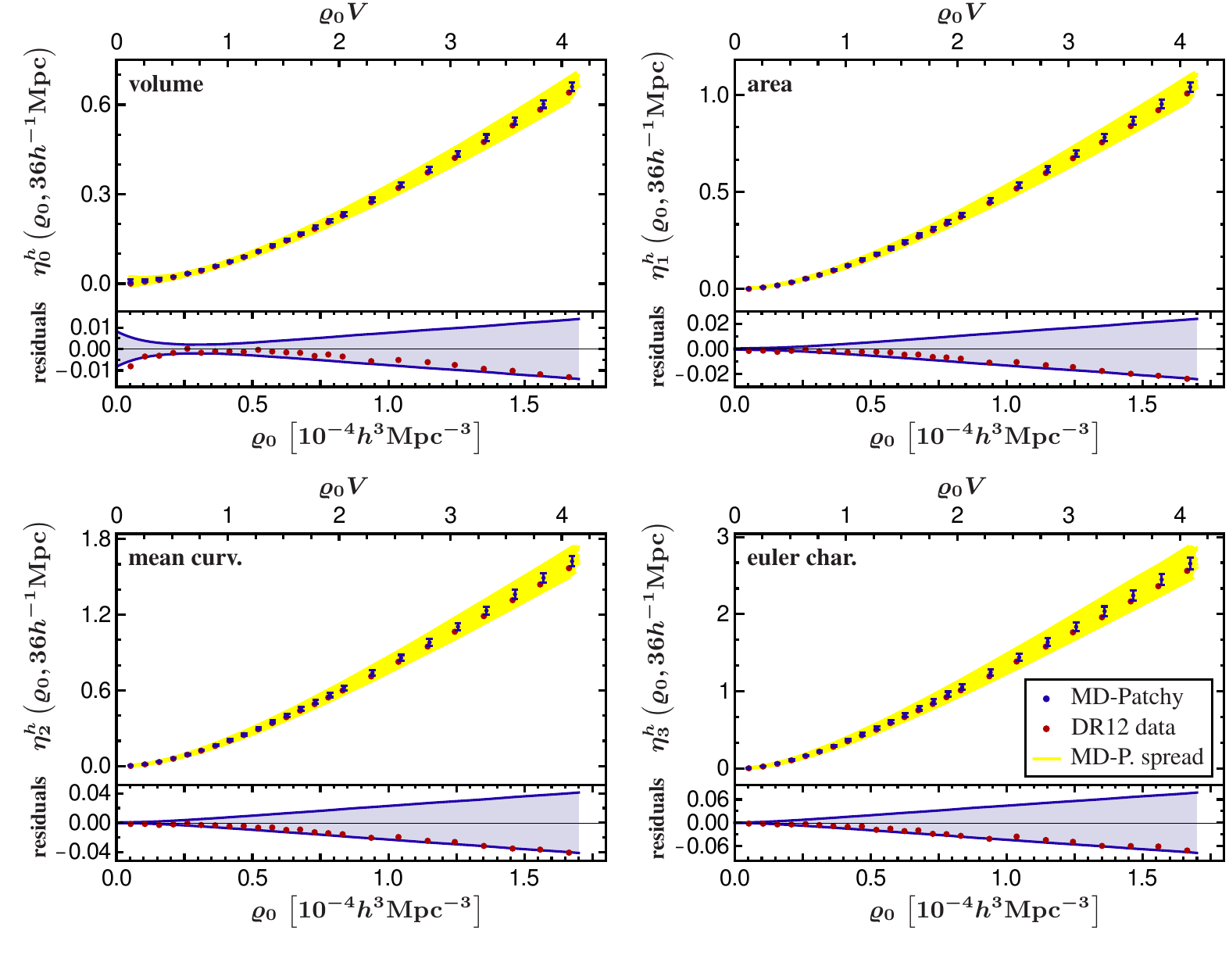}
\end{centering}

\caption{Comparison of the higher-order ($3$-pt. and higher) part $\eta_{\mu}^{h}$
of the MD-Patchy mocks (blue with error bars) with the data (red points)
for a ball diameter of $36\protect\hMpc$. This is basically obtained
by subtracting the straight two-point lines of Fig.~\ref{fig:ModMinkDensDep}
from the curved $\eta_{\mu}$ lines. The error bars are $1-\sigma$
and the yellow bands show the full spread of the 900 mocks. The lower
part of the plots give the residuals of the galaxy data w.r.t. the
mean of the mocks. For all four functionals the data happens to be
around $1-\sigma$ below the mocks.\label{fig:ModMinkSubtract}}
\end{figure*}

This becomes even more apparent in the plots of Fig.~\ref{fig:ModMinkSubtract}.
\begin{table}
\begin{centering}
\begin{tabular}{cccccc}   \hline\hline \vspace{-0.3cm} \\ & \multicolumn{2}{c}{MD-P. $\eta^{h}_{\mu}(\varrho_{0})$} \hspace{-0.3cm} & &  \multicolumn{2}{c}{MD-P. $\eta^{h}_{\mu}(\varrho_{0})$} \hspace{-0.3cm} \\ \vspace{-0.3cm} \\   & $\chi^2$ & $\sigma_G$ & & $\chi^2$ & $\sigma_{G}$ \\  \noalign{\smallskip} \hline \noalign{\smallskip} $\eta_0$ & 10300 & 97.3 & & 290 & 5.50   \\  $\eta_1$ & 11600 & 104 & & 282 & 5.22   \\  $\eta_2$ & 12300 & 107 & & 221 & 2.76   \\  $\eta_3$ & 12500 & 108 & & 252 & 4.04   \\ \noalign{\smallskip}\hline \end{tabular}
\par\end{centering}

\caption{First column: $\chi^{2}$ values for the deviation of the higher-order-only
$\eta_{\mu}^{h}$ of the MD-Patchy mocks in Fig.~\ref{fig:ModMinkSubtract}
from the zero line. The value includes all $171$ grid points of Fig.~\ref{fig:DensityGrid}
between $18\protect\hMpc$ and $66\protect\hMpc$ in diameter. The
values indicate a significant contribution of higher-order correlations.\protect\\Second
column: Combined $\chi^{2}$ values for the deviation of the SDSS
data points in Fig.~\ref{fig:ModMinkSubtract} from the MD-Patchy
mocks for the same $171$ grid points as the first column. \label{tab:ChiSqRadComb}}
\end{table}
For these we use the integrated two-point correlations and subtract
them off the corresponding $\eta_{\mu}$ for each mock individually.
From these subtracted quantities $\eta_{\mu}^{h}$ (Eq.~(\ref{eq:DefModMinkHigher}))
we calculate the mean of the mocks and the covariances of the density
points. The plots of the residuals show, that the higher-order contributions
to the modified MFs are quite well described by the structure in the
MD-Patchy mocks. The data happens to lie close to the $1-\sigma$
line of the fluctuations between the $900$ mocks for all four MFs.
It is quite encouraging that the mocks that were constructed to match
the two- and three-point correlations only also capture what is going
on in the higher orders quite well.

As the data points in both Fig.~\ref{fig:ModMinkDensDep} and Fig.~\ref{fig:ModMinkSubtract}
are strongly correlated, we need to take into account the full covariance
matrix in order to estimate the significance of the remaining difference.
The results of this calculation are shown in Tab.~\ref{tab:ChiSqMockDat}.
For the curves of Fig.~\ref{fig:ModMinkDensDep} and Fig.~\ref{fig:ModMinkSubtract}
with $24$ density points, we first evaluate the $\chi^{2}$ values
and then convert their $p$-values into those of a two sided deviation
for a Gaussian, which we express as multiples of one standard deviation.
We inflate the error bars by multiplying the inverse covariance matrix
by a factor of $\left(1-N_{pts}/N_{mocks}\right)$, in order to account
for the finite number of mocks we use for estimating the covariance
matrix \citep{2014MNRAS.439.2531P}. We find that the MD-Patchy
mocks are a much better match of the data than the QPM mocks which
are about $50\sigma$ away from the data in this case. For the higher
components only, i.e. the central column of Tab.~\ref{tab:ChiSqMockDat},
the agreement is even better and does not exceed $2\sigma$ for any
of the functionals. So also quantitatively with covariances included
the impression from the residual plots of Fig.~\ref{fig:ModMinkSubtract}
that the typical deviation is around $1\sigma$ is correct. 

\begin{figure*}
\begin{centering}
\includegraphics[width=0.9\textwidth]{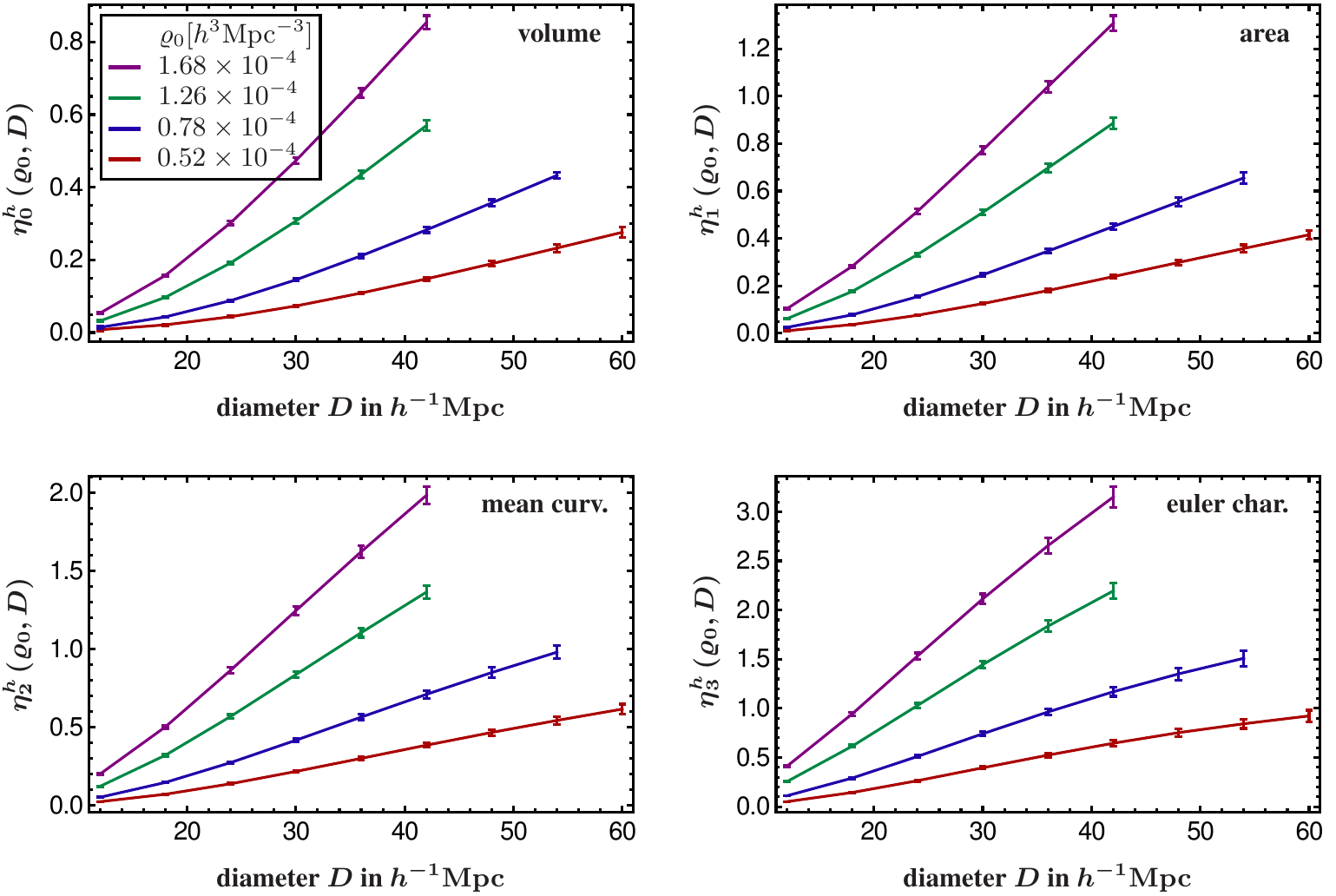}
\end{centering}

\caption{Evolution of the higher-order only part $\eta_{\mu}^{h}\left(\varrho_{0},R\right)$
of Eq.~(\ref{eq:DefModMinkHigher}) with ball diameter $2R$ for
four different densities. The full density dependence for a single
diameter of $2R=36\protect\hMpc$ was shown in Fig.~\ref{fig:ModMinkSubtract}.
The higher-order signal increases with scale and density. \label{fig:ModMinkSubtractRad}}
\end{figure*}

When we combine all the information of the relevant $171$ points
in our sampling grid of Fig.~\ref{fig:DensityGrid}, i.e. ball diameters
from $18\hMpc$ to $66\hMpc$ and their complete density dependence,
we find in Tab.~\ref{tab:ChiSqRadComb} that the deviation of the
higher-order MD-Patchy modified MFs is at most $5.5\sigma$. For some
diameters as in Tab.~\ref{tab:ChiSqMockDat} the agreement is nearly
perfect. This is very encouraging, as the mocks have only been generated
to match the two- and three-pt. function. Given the combined values
of Tab.~\ref{tab:ChiSqRadComb}, it is clear that there is still
some room for improvement in the construction of realistic mocks to
match the higher orders even better. 

Regarding the deviation of the $\eta_{\mu}^{h}$ from the case of
vanishing higher-order correlations (the zero line in Fig.~\ref{fig:ModMinkSubtract}),
the conclusion is clear. There is strong evidence for higher-order
correlations in the mocks and the data as seen by the MFs. Even though
this is not surprising, as at least the three-point function of the
SDSS data is well known (see e.g. \citet{2015arXiv151202231S} and
\citet{2009MNRAS.399..801G,2011ApJ...737...97M}), it is still interesting
to see the statistical power that we have at hand for testing models
with different higher-order correlations. 

For all of the quantities of Tabs. \ref{tab:ChiSqMockDat} and \ref{tab:ChiSqRadComb},
one should keep in mind that the different MFs are not independent.
So in order to derive one single value that would quantify the mismatch
between mocks and data, one would also here have to take the covariances
into account. For a single radius this is possible. For the full data
set, we are limited by the number of mocks that we analyzed as an
estimate of the $684\times684$ covariance matrix would be noisy.

Finally, let us illustrate the scale dependence of the higher order
contributions. Fig.~\ref{fig:ModMinkSubtractRad} shows the deviation
of the full $\eta_{\mu}$ (\ref{eq:MinkCorrCon}) from the two-point
part (\ref{eq:MinkGaussExp}), i.e. $\eta_{\mu}^{h}$ of (\ref{eq:DefModMinkHigher}).
So it is the same quantities as in Fig.~\ref{fig:ModMinkSubtract},
but now as a function of ball diameter for four fixed densities $\varrho_{0}\in\left\{ 0.52,0.78,1.26,1.68\right\} \times10^{-4}\hMpc$.
The higher the density, the larger is the deviation from the two-point
part, because the higher-order contributions play a more important
role in the $\varrho_{0}$-power series. In the radial direction,
the functionals probe the higher-order correlations at different scales.
For the two-point part we have seen in Fig.~\ref{fig:IntegrationWindow}
that the integration kernel picks up correlations at larger scales
for increasing functional index $\mu$. This is most probably also
the case for the higher-order correlations (we explicitly checked
it for the three-point integrals) and so $\eta_{3}^{h}$ probes the
largest scales that we accessed in this analysis. The decreasing slope
of the radial dependence reflects a similar behavior in the three
point integral of $\xi_{3}$. For even larger scales, the latter integral
becomes negative and the difference to the two-point case decreases
again.

\subsection{Integrated n-point correlation functions\label{sub:SeriesCoefficients}}

Having found a significant contribution of higher-order correlations
to the MFs, we now want to study which orders actually contribute.
Using the relation in Eq.~(\ref{eq:MinkCorrCon}) we can decompose
the functionals in different contributions from higher orders by fitting
a polynomial of order $n$ to the density dependence of the $\eta_{\mu}$
as described in Sec.~\ref{sub:Density-dependence}. This decomposition
is not necessary for using the higher-order MF information in the
comparison of the model to the data, as we have shown in the previous
section, but it illustrates the convergence properties of the $\eta_{\mu}$
series. 

\begin{figure}
\begin{centering}
\includegraphics[width=0.45\textwidth]{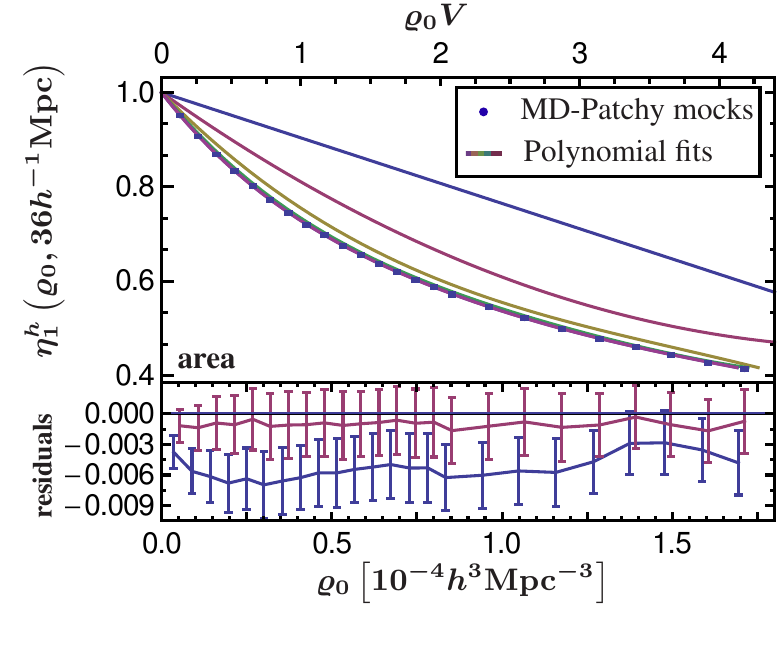}
\end{centering}
\caption{Illustration of the extraction of the fit coefficients $c_{\mu,n}$
of Eq.~(\ref{eq:MinkIntegr}). To the $\varrho_{0}$-dependence of
$\eta_{1}$ at a diameter of $36\protect\hMpc$ we fit polynomials
in $\varrho_{0}$ of increasing order taking the full correlations
of the points into account and extract the $c_{\mu}$ from the best
fit coefficients. The linear to third order polynomials in the large
panel give a terrible fit already visually. The fourth and fifth order
in the ``residual'' part of the plot indicate convergence at 5.-order.
The corresponding $\chi^{2}$ values can be found in Tab.~\ref{tab:HigherFitChi}
and confirm this fact.\label{fig:HigherFit}}
\end{figure}

In Fig.~\ref{fig:HigherFit} we evaluate the convergence of the fit
for increasing polynomial order $n$.
The deviations of the fitted model from the data are quantified in
Tab.~\ref{tab:HigherFitChi}.

From both presentations it is clear that the linear term that only
includes the two-point coefficient $c_{\mu,2}$ is a very bad fit
to the density evolution. The table indicates that only after the
inclusion of the fifth term, which corresponds to integrals over the
six-point function, the polynomial fit captures the shape of the modified
MFs well. However, this only means that we have some sensitivity of
the MFs of the mocks to contributions from up to the six-point function.
It does not mean that these contributions are also significantly different
from the ones in the data. In fact, as we have found in Sec.~\ref{sub:HighSubtract}
the higher-order contributions of the mocks and the data actually
agree relatively well.

\begin{figure*}
\begin{centering}
\includegraphics[width=0.9\textwidth]{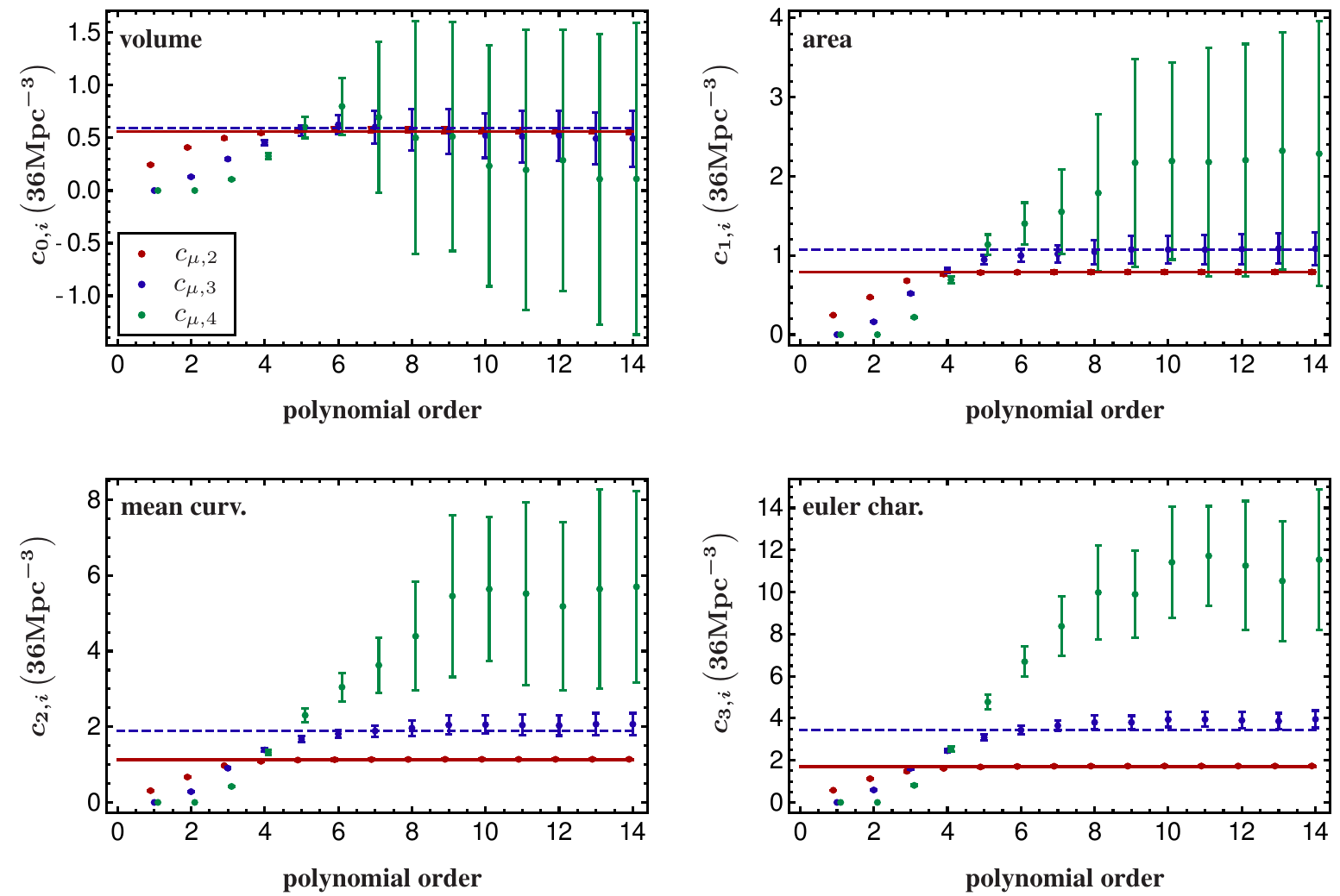}
\end{centering}
\caption{Evolution of the best fit value of the first three coefficients $c_{\mu,n+1}$
of Eq.~(\ref{eq:MinkIntegr}) with fit order $n$ for a ball diameter
of $36\protect\hMpc$. $c_{\mu,2}<c_{\mu,3}<c_{\mu,4}$ for large
$n$. The solid (red) lines are the coefficients calculated from the
two-point integrals of our measurement of the correlation function.
The dashed (blue) lines are the predictions from the integral over
the model three-point correlation function of \protect\citet{2016arXiv160703109S}
with $b_{1}=2.19$, $b_{2}=0$ and $b_{t}=-0.34$.\label{fig:EvWithFitOrder}}
\end{figure*}

In order to evaluate how strongly the coefficients $c_{\mu,n}$ of
Eq.~(\ref{eq:PowerSeriesDecomp}) are biased by truncating the power
series at a finite $n$, we plot in Fig.~\ref{fig:EvWithFitOrder}
the dependence of the mean of the best fit values of $c_{1,2}$, $c_{1,3}$
and $c_{1,4}$ of the $900$ MD-Patchy mocks on the polynomial order
$n$ for a fit at a diameter of $36\hMpc$. As high values of $n$
also increase the chance of picking up spurious fluctuations, we decided
to put a prior on the higher order coefficients. To this end we fit
a three-parameter Pad{\'e} model to the density evolution of the
mean MFs extracted from $100$ periodic MD-Patchy cubes. Taylor expanding
this model provides some model for the higher-order coefficients.
We choose a Gaussian prior around the so obtained values with a standard
deviation corresponding to the value of the mean $G\left(\mu,\sigma=\mu\right)$.
We only apply this prior to coefficients beyond the four-point function
coefficient $c_{\mu,4}$. The first two coefficients $c_{\mu,2}$
and $c_{\mu,3}$ are already well measured without any prior, and
for $c_{\mu,4}$ it is sufficient to regularize the higher orders.
Fig.~\ref{fig:EvWithFitOrder} shows that the value of the first
coefficient $c_{\mu,2}$ is independent of $n$ as soon as $n$ is
high enough that we have a good fit, i.e. $n=5$ from Fig.~\ref{fig:HigherFit}.
The second coefficient $c_{\mu,3}$ is also largely independent of
fit order, but needs a higher fit order $n$ than $c_{\mu,2}$ to
be unbiased. For $c_{\mu,4}$ there is still some remaining uncertainty
from the fit order, which is, however, still much smaller than the
error from the scatter of the mocks.

\begin{figure}
\begin{centering}
\includegraphics[width=0.45\textwidth]{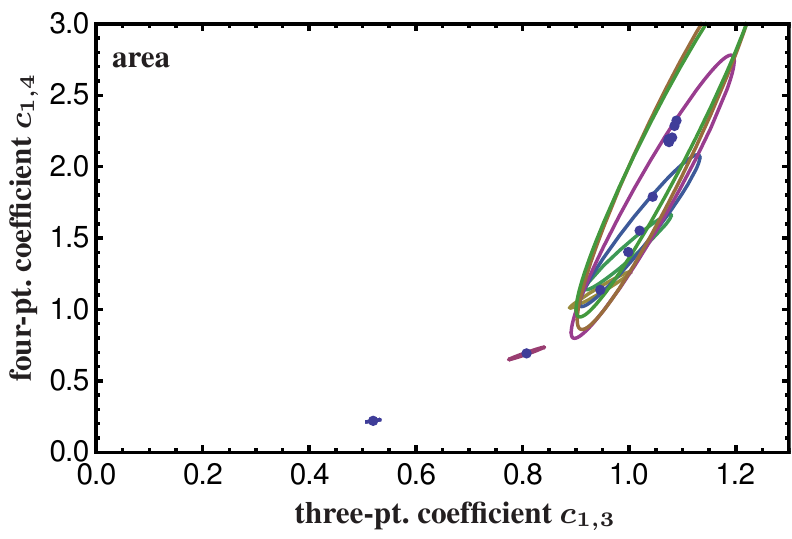}
\end{centering}
\caption{Evolution of the best fit coefficients $c_{1,3}$ and $c_{1,4}$ of
the fit to $\eta_{1}\left(\varrho_{0}\right)$ of Fig.~\ref{fig:HigherFit}
for a diameter of $36\protect\hMpc$. The leftmost point is for a
third order polynomial and the order is increasing from left to right
with a maximum order of $14$. The error bars indicate $1-\sigma$
intervals. Starting from the ninth order, error bars include all higher
order points.\label{fig:ErrorEllipse}}
\end{figure}

Fig.~\ref{fig:ErrorEllipse} shows evolution of the mean value together
with the $1-\sigma$ contours of the parameters $c_{1,3}$ and $c_{1,4}$
for different $n$ starting at $n=3$, to include also the covariances
between the parameters.
Below the order $5$ that provides the first good fit in Tab.~\ref{tab:HigherFitChi},
the best fit values are strongly dependent on fit order. For orders
starting at $n=9$ the shift becomes much smaller than the scatter
of the mocks. For the following plots, we will use polynomials of
order $9$ for the $\eta_{1}$ fit at $D=36\hMpc$. To determine the
fit order for the radii with a smaller number of density points, i.e.
beyond $42\hMpc$, we constructed and analyzed test data. These were
created from a Pad{\'e} model fit to the cut sky mocks and provide
a model with known coefficients. Then we applied the fitting procedure
to these test data and found that for high enough $n$, the coefficients
we put in were recovered within the errors estimated by the fluctuations
from the mocks. From the results of these test data fits, we determined
for every diameter the minimum $n$ needed. For the larger diameters
this value is lower due to the decreasing contribution of higher orders.
The reason for this decrease is the steep scale dependence for the
higher-order coefficients, an example of which we will see in Fig.~\ref{fig:ThreePtCoeffs}
for the three-point term (see also Fig.~14 of \citet{2014MNRAS.443..241W}
for the scale dependence of the coefficients of $\eta_{0}$ for a
Log-Normal model).

The decomposition into $n$-point components also allows us to test
to which extent the deviation of the data from the mocks in Fig.~\ref{fig:ModMinkDensDep}
could be simply the result of a misnormalization of the power spectrum.
The green line in Fig.~\ref{fig:ModMinkDensDep} therefore shows
the mean of the MD-Patchy mocks for a reduction in the power of $5\%$.
It has been obtained by taking the coefficients of the $6$-th order
fit of Fig.~\ref{fig:HigherFit} and multiplying the integrated correlation
functions with the power of the reduction factor that corresponds
to the number of power spectra involved, i.e. $0.95^{n}$ for $\xi_{n+1}$.
Of course this is only a very simplified model, given that the fit
is not unbiased, especially for higher orders, but it should still
provide a useful approximation. From the result in Fig.~\ref{fig:HigherFit}
we see that at least approximately it removes the offset of the data
from the mocks and makes them agree even better. This seems to be
consistent with the analysis of the power spectrum in \citet{2016arXiv160703149B}.
Our redshift range roughly corresponds to their second redshift bin,
for which their Fig.~1 indicates that the data also is slightly lower
than the mean of the MD-Patchy mocks.

\begin{table}
\begin{centering}
\begin{tabular}{cccc}  \hline\hline \vspace{-0.3cm} \\ fit order & coeff. & $\chi^2$ & p-value \\  \noalign{\smallskip} \hline \noalign{\smallskip}  $1$ & $c_{1,2}$ & 7859 & $10^{-1676}$  \\   $2$ & $c_{1,3}$ & 1509 & $10^{-306}$ \\ $3$ & $c_{1,4}$ & 306 & $10^{-52}$   \\   $4$ & $c_{1,5}$ & 47 & $6.2\cdot 10^{-4}$  \\  $5$ & $c_{1,6}$ & 11.6 & $9.7\cdot 10^{-2}$ \\ \noalign{\smallskip}\hline\end{tabular}
\par\end{centering}

\caption{$\chi^{2}$ values for the deviation of the best fit polynomial from
the $\varrho_{0}$-dependence of $\eta_{1}$ for the mean of $900$
MD-Patchy mocks at a ball diameter of $36\protect\hMpc$. The corresponding
curves are shown in Fig.~\ref{fig:HigherFit}. There are $24$ points
to fit. Only after adding the fifth coefficient, corresponding to
the $6$-point function integral, to the fit, the $\chi^{2}$ values
become acceptable.\label{tab:HigherFitChi}}
\end{table}

\begin{figure*}
\begin{centering}
\includegraphics[width=0.87\textwidth]{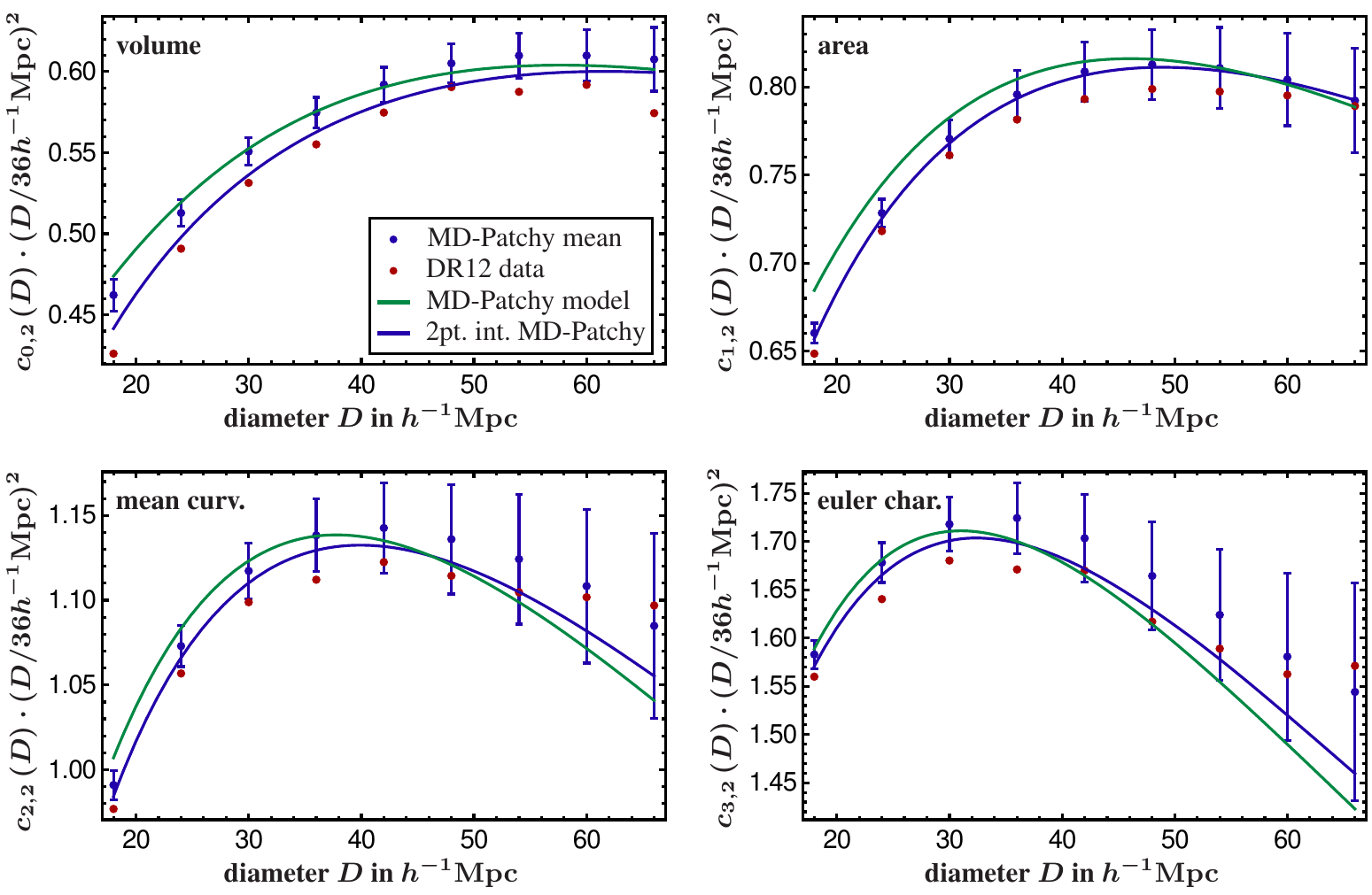}
\end{centering}
\caption{Linear term of the $\varrho_{0}$-series Eq.~(\ref{eq:PowerSeriesDecomp})
for the MD-Patchy mocks and the DR12 galaxy data. The linear coefficient
corresponds to the integrated two-point correlation function of Eq.~(\ref{eq:MinkIntegr}).
It is obtained by fitting a polynomial to the density dependence of
the $\eta_{\mu}$ for different ball diameters. The fit is illustrated
in Fig.~\ref{fig:HigherFit}. The blue solid line is the corresponding
integral determined from an independent measurement of the two-point
correlation function. The green line is the integral of the non-linear
MD-Patchy power spectrum at $z=0.50$ with a linear bias of $b_{1}=2.19$.\label{fig:TwoPtCoeffs}}
\end{figure*}

\begin{figure*}
\begin{centering}
\includegraphics[width=0.87\textwidth]{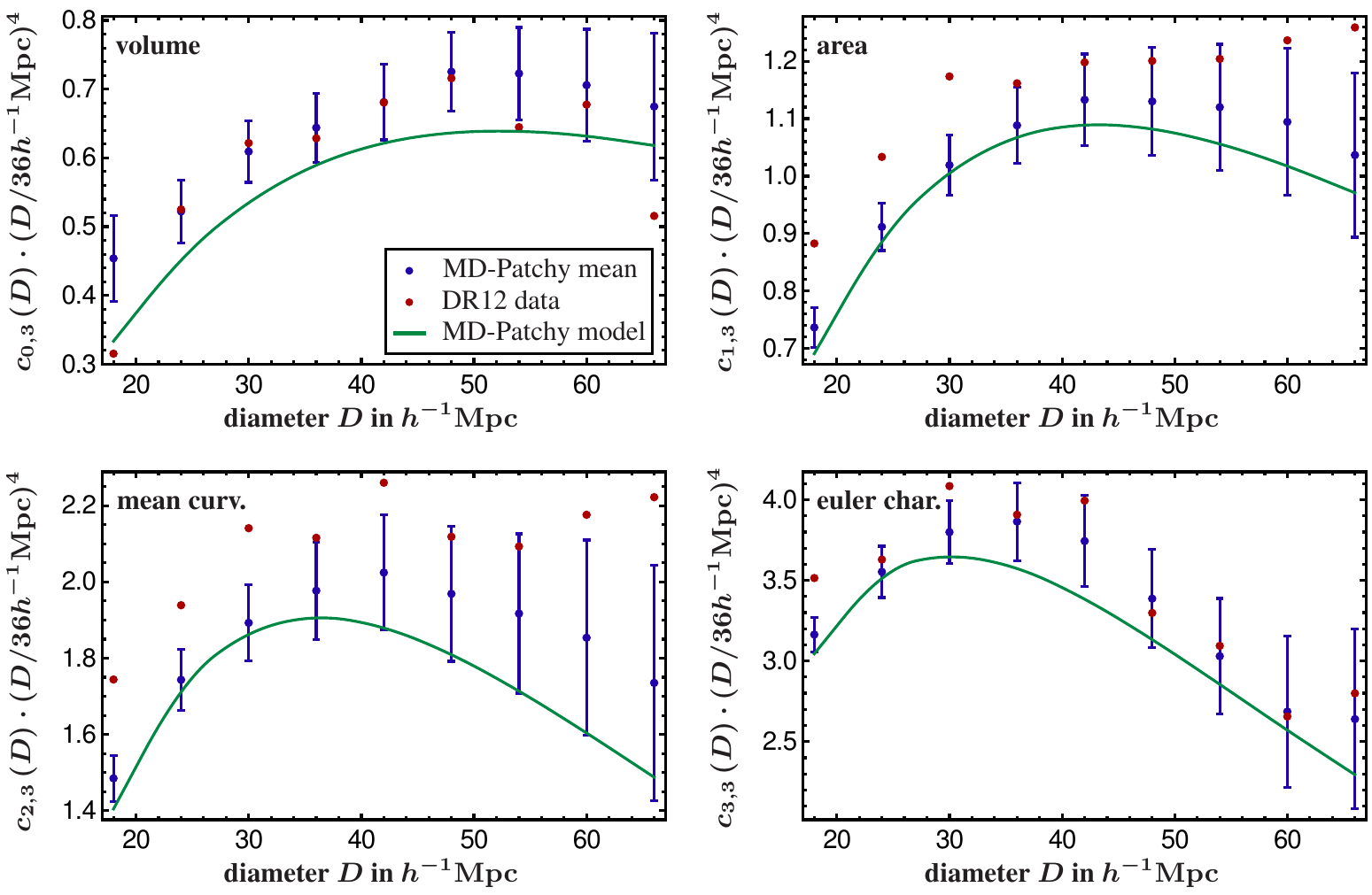}
\end{centering}
\caption{Quadratic term of the $\varrho_{0}$-series Eq.~(\ref{eq:PowerSeriesDecomp})
for the MD-Patchy mocks and the DR12 galaxy data. This quadratic coefficient
corresponds to the integrated three-point correlation function of
Eq.~(\ref{eq:MinkIntegr}). The result is from the same fit as Fig.~\ref{fig:TwoPtCoeffs}.
The solid line is the integral over the model three-point correlation
function of \protect\citet{2016arXiv160703109S} with $b_{1}=2.19$ as above,
$b_{2}=0$ and $b_{t}=-0.34$. \label{fig:ThreePtCoeffs}}
\end{figure*}

For the two cases $c_{\mu,2}$ and $c_{\mu,3}$ where the convergence
in Fig.~\ref{fig:EvWithFitOrder} is convincing, we compare in Figs.~\ref{fig:TwoPtCoeffs}
and \ref{fig:ThreePtCoeffs} the extracted coefficients to reference
models. Fig.~\ref{fig:TwoPtCoeffs} shows the two-point coefficients
$c_{\mu,2}$. For those, $c_{0,2}$ coincides with $\sigma^{2}\left(R\right)$
(see Eq.~(\ref{eq:V0-sigma})) and $c_{3,2}$ corresponds to the
volume averaged correlation function (see Eq.~(\ref{eq:V3-xi})).
In addition to the average of the coefficients extracted from the
MD-Patchy mocks, we show the coefficients of the data and the expected
coefficients derived from integrals over the average two-point correlation
function that we measured independently. The overall agreement of
the two independent determinations is reassuring. It provides a confirmation
that the fit to $\eta_{\mu}\left(\varrho_{0}\right)$ can extract
individual terms from the power series (\ref{eq:MinkCorrCon}). There
is, however, for some of the functionals a relatively large offset.
It is not clear if this offset is due to residual bias in the fitting
method, or if our method to estimate the correlation function of a
mock does not include the boundaries in the same way than the MFs.

Fig.~\ref{fig:ThreePtCoeffs} then shows the $c_{\mu,3}$.
The connection to other known quantities is not as apparent as in
the two-point case and we did not obtain an independent estimate of
the three-point correlation integrals. What is shown in Fig.~\ref{fig:ThreePtCoeffs}
instead is the three-point function model of \citet{2016arXiv160703109S}.
It is based on leading-order perturbation theory and includes redshift
space distortions. For the plots in Fig.~\ref{fig:ThreePtCoeffs}
we fixed the three bias parameters to $b_{1}=2.19$, $b_{2}=0$ and
$b_{t}=-0.34$. The $b_{1}$-value is the one we used in Fig.~\ref{fig:TwoPtCoeffs}
and the $b_{t}$-value is obtained from Eq.~(23) in \citet{2016arXiv160703109S}
and based on local Lagrangian biasing \citep{2012PhRvD..86h3540B,2012PhRvD..85h3509C}.
We did not attempt to fit for $b_{2}$ and so the lines in Fig.~\ref{fig:ThreePtCoeffs}
do not contain any free parameter. Given that, the match of the model
with the extracted coefficients is encouraging.

\subsection{The occurrence of large voids\label{sub:Voidfraction}}

In addition to being related to the hierarchy of $n$-point correlation
functions, the MFs also contain information about the cosmic web.
For example, the partial functionals probe the local morphology around
the galaxy sites, and for an isodensity contour smoothing ``shapefinders''
have been defined to quantify the amount of filamentary vs. planarity
of the local density field \citep{1998ApJ...495L...5S,1999ApJ...526..568S,phdSchmalzing}. 

For the global germ-grain MF densities that we study here, there are
different quantities that can be extracted. The Euler characteristic
$\chi=v_{3}V$, for example, indicates for large balls how many extreme
voids (i.e. regions that are further away from any galaxy than $2R_{max}$)
there are in the galaxy distribution. The area functional $v_{1}$
gives an idea of their area and $v_{0}$ their volume. More generally,
the comparison of the void fraction $1-v_{0}$ to the one of a Poisson
distribution of the same density indicates how inhomogeneously these
void regions are distributed.

\begin{figure}
\begin{centering}
\includegraphics[width=0.45\textwidth]{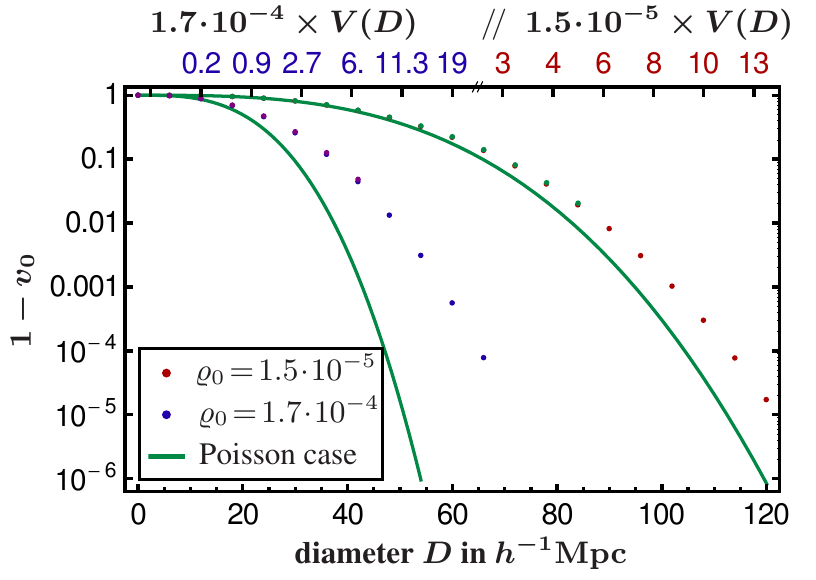}
\end{centering}
\caption{Fraction of unfilled volume for two different sample densities as
a function of the ball radius. The top scale gives the corresponding
$\varrho V\left(D\right)$ for a ball of diameter $D$, where the
first values correspond to the lower (blue) points and the values
restarting at $3$ to the upper (red) points. The densities are $\varrho_{1}=1.68\times10^{-4}\protect\hDens$
and $\varrho_{24}=1.54\times10^{-5}\protect\hDens$. Especially for
the higher density the solid Poisson lines lie way below the values
for the data, so the structure leads to a significantly higher void
fraction than we would expect in an unclustered field.\label{fig:FreeVolumeFraction}}
\end{figure}

In Fig.~\ref{fig:FreeVolumeFraction}, we therefore show $1-v_{0}$
as a function of the ball diameter and compare it to the corresponding
Poisson case.
Especially for the higher density of $\varrho_{1}=1.68\times10^{-4}\hDens$
the difference is quite remarkable. For a diameter of $42\hMpc$,
the volume is a factor of $\approx30$ larger than in the Poisson
case. For the most extreme case of $66\hMpc$, it is nearly $10^{7}$.
In this latter case the expected number of galaxies in a ball of diameter
$66\hMpc$ would be $25$. With the standard Poisson fluctuation one
would expect a fraction of $\approx10^{-11}$ to be uncovered. Including
the observed structure however, there is $\approx8\times10^{-5}$
of the $1.38\hVolG$ in our analyzed region that remain uncovered.

Of course one has to make sure that this conclusion is not significantly
affected by the holes in the survey. We tested this using the periodic
cubic MD-Patchy mocks and plot the resulting void fractions as purple
and green points in Fig.~\ref{fig:FreeVolumeFraction}. For the region
that we evaluated, the cubic mocks for the difference is small, below
$4\%$ for $\varrho_{1}$ and below $9\%$ for $\varrho_{24}$. However,
for the more extreme diameters the deviations will grow, but we do
not expect that this affects the general picture.

\section{Conclusion}\label{sec:Conclusion}

In this paper, we perform a new analysis that combines Minkowski Functionals
with the standard two-point correlation function to measure the amount
of non-Gaussianity of the density field. We find in Sec.~\ref{sub:HighSubtract}
that the significance of the higher-order contribution to the MFs
is of the order of $100\sigma$ (see Tab.~\ref{tab:ChiSqRadComb}
and Fig.~\ref{fig:ModMinkSubtract}). This demonstrates that this
part of the MFs can be well enough measured to give meaningful constraints
on the higher-order properties of the galaxy distribution.

Given this constraining power it is reassuring that we find from Fig.~\ref{fig:ModMinkSubtract}
and Tab.~\ref{tab:ChiSqMockDat} that the MD-Patchy mocks, that were
constructed to match the two- and three-point properties of the measured
galaxy distribution, match fairly well to the higher-order MFs. This
indicates that as far as the higher-orders accessible by the MFs are
concerned, the mock galaxy catalogs are quite good. There is still
some room for improvement as the combined deviation is as large as
$5\sigma$ for some functionals (see Tab.~\ref{tab:ChiSqRadComb}),
but this might be addressed by a $5\%$ shift in the normalization
of the mock power spectrum (see Fig.~\ref{fig:ModMinkDensDep}).

On the technical side, our careful boundary treatment minimizes the
effects of the survey geometry on our measurement of the modified
MFs (see Fig.~\ref{fig:MaskCorrection}). The corrected values lie
within the error bars of a boundaryless periodic box.

For the QPM mocks, we find a much larger deviation (beyond $10\sigma$)
from the data than for MD-Patchy (see Tab.~\ref{tab:ChiSqMockDat}).
As in this case already the two-point correlations are overnormalized
in the range of scales that we are interested in (see Fig.~\ref{fig:MeanCorrFun}),
a reduction of the amplitude of the initial Gaussian field could also
help in this case.

We show that in the the higher-order part of the MFs calculated for
the SDSS DR12 CMASS galaxies contains at least information up to the
integrated $6$-point function (see Tab.~\ref{tab:HigherFitChi}).
This underlines the power of the MFs to bring condensed higher-order
information into a simple functional form.

\subsubsection*{Acknowledgements}

AW thanks Antonio Cuesta for providing support with the QPM correlation
functions. We thank Graziano Rossi and Thomas Buchert for valuable comments on the manuscript.
The work of AW was supported by the German research organization
DFG, Grant No. WI 4501/1-1. 
DJE is supported by U.S. Department of Energy
grant DE-SC0013718 and as a Simons Foundation Investigator.

Funding for SDSS-III has been provided by the Alfred P. Sloan Foundation,
the Participating Institutions, the National Science Foundation, and
the U.S. Department of Energy Office of Science. The SDSS-III web
site is http://www.sdss3.org/.

SDSS-III is managed by the Astrophysical Research Consortium for the
Participating Institutions of the SDSS-III Collaboration including
the University of Arizona, the Brazilian Participation Group, Brookhaven
National Laboratory, Carnegie Mellon University, University of Florida,
the French Participation Group, the German Participation Group, Harvard
University, the Instituto de Astrofisica de Canarias, the Michigan
State/Notre Dame/JINA Participation Group, Johns Hopkins University,
Lawrence Berkeley National Laboratory, Max Planck Institute for Astrophysics,
Max Planck Institute for Extraterrestrial Physics, New Mexico State
University, New York University, Ohio State University, Pennsylvania
State University, University of Portsmouth, Princeton University,
the Spanish Participation Group, University of Tokyo, University of
Utah, Vanderbilt University, University of Virginia, University of
Washington, and Yale University.


\begin{thebibliography}{}

\bibitem[\protect\citeauthoryear{{Aihara}, {Allende Prieto}, {An}, {Anderson},
  {Aubourg}, {Balbinot}, {Beers}, {Berlind}, {Bickerton}, {Bizyaev}, {Blanton},
  {Bochanski} \& {et al.}}{{Aihara} et~al.}{2011}]{2011ApJS..193...29A}
{Aihara} H. et~al., 2011, \apjs, 193, 29, \eprintc{1101.1559},
  \adsurl{http://adsabs.harvard.edu/abs/2011ApJS..193...29A}

\bibitem[\protect\citeauthoryear{{Alam}, {Albareti}, {Allende Prieto},
  {Anders}, {Anderson}, {Anderton}, {Andrews}, {Armengaud}, {Aubourg}, {Bailey}
  \& et al.}{{Alam} et~al.}{2015}]{2015ApJS..219...12A}
{Alam} S. et~al., 2015, \apjs, 219, 12, \eprintc{1501.00963},
  \adsurl{http://adsabs.harvard.edu/abs/2015ApJS..219...12A}

\bibitem[\protect\citeauthoryear{{Alam}, {Ata}, {Bailey}, {Beutler}, {Bizyaev},
  {Blazek}, {Bolton}, {Brownstein}, {Burden}, {Chuang}, {Comparat} \&
  {Cuesta}}{{Alam} et~al.}{2016}]{2016arXiv160703155A}
{Alam} S. et~al., 2016, ArXiv e-prints, \eprintc{1607.03155},
  \adsurl{http://adsabs.harvard.edu/abs/2016arXiv160703155A}

\bibitem[\protect\citeauthoryear{{Aragon-Calvo}, {Shandarin} \&
  {Szalay}}{{Aragon-Calvo} et~al.}{2010}]{2010arXiv1006.4178A}
{Aragon-Calvo} M.~A., {Shandarin} S.~F., {Szalay} A. 2010, ArXiv e-prints,
  \eprintc{1006.4178},
  \adsurl{http://adsabs.harvard.edu/abs/2010arXiv1006.4178A}

\bibitem[\protect\citeauthoryear{{Baldauf}, {Seljak}, {Desjacques} \&
  {McDonald}}{{Baldauf} et~al.}{2012}]{2012PhRvD..86h3540B}
{Baldauf} T., {Seljak} U., {Desjacques} V., {McDonald} P. 2012, \prd, 86,
  083540, \eprintc{1201.4827},
  \adsurl{http://adsabs.harvard.edu/abs/2012PhRvD..86h3540B}

\bibitem[\protect\citeauthoryear{{Beutler}, {Seo}, {Ross}, {McDonald}, {Saito},
  {Bolton}, {Brownstein}, {Chuang}, {Cuesta} \& {Eisenstein}}{{Beutler}
  et~al.}{2016}]{2016arXiv160703149B}
{Beutler} F. et~al., 2016, ArXiv e-prints, \eprintc{1607.03149},
  \adsurl{http://adsabs.harvard.edu/abs/2016arXiv160703149B}

\bibitem[\protect\citeauthoryear{{Blake}, {James} \& {Poole}}{{Blake}
  et~al.}{2014}]{2014MNRAS.437.2488B}
{Blake} C., {James} J.~B., {Poole} G.~B. 2014, MNRAS, 437, 2488,
  \eprintc{1310.6810},
  \adsurl{http://adsabs.harvard.edu/abs/2014MNRAS.437.2488B}

\bibitem[\protect\citeauthoryear{{Bolton}, {Schlegel}, {Aubourg}, {Bailey},
  {Bhardwaj} \& {et al.}}{{Bolton} et~al.}{2012}]{2012AJ....144..144B}
{Bolton} A.~S. et~al., 2012, AJ, 144, 144, \eprintc{1207.7326},
  \adsurl{http://adsabs.harvard.edu/abs/2012AJ....144..144B}

\bibitem[\protect\citeauthoryear{{Buchert}}{{Buchert}}{1995}]{1995lssu.conf..156B}
{Buchert} T., 1995, in {M{\"u}cket} J.~P.,  {Gottl{\"o}ber} S.,   {M{\"u}ller}
  V.,  eds, Large Scale Structure in the Universe {Robust Morphological
  Measures for Large Scale Structure}.
p.~156, \eprintc{astro-ph/9412061},
  \adsurl{http://adsabs.harvard.edu/abs/1995lssu.conf..156B}

\bibitem[\protect\citeauthoryear{{Chan}, {Scoccimarro} \& {Sheth}}{{Chan}
  et~al.}{2012}]{2012PhRvD..85h3509C}
{Chan} K.~C., {Scoccimarro} R., {Sheth} R.~K. 2012, \prd, 85, 083509,
  \eprintc{1201.3614},
  \adsurl{http://adsabs.harvard.edu/abs/2012PhRvD..85h3509C}

\bibitem[\protect\citeauthoryear{{Choi}, {Kim}, {Rossi}, {Kim} \& {Lee}}{{Choi}
  et~al.}{2013}]{2013ApJS..209...19C}
{Choi} Y.-Y. et~al., 2013, \apjs, 209, 19, \eprintc{1309.4381},
  \adsurl{http://adsabs.harvard.edu/abs/2013ApJS..209...19C}

\bibitem[\protect\citeauthoryear{{Codis}, {Pichon}, {Pogosyan}, {Bernardeau} \&
  {Matsubara}}{{Codis} et~al.}{2013}]{2013MNRAS.435..531C}
{Codis} S. et~al., 2013, MNRAS, 435, 531, \eprintc{1305.7402},
  \adsurl{http://adsabs.harvard.edu/abs/2013MNRAS.435..531C}

\bibitem[\protect\citeauthoryear{{Cuesta}, {Vargas-Maga{\~n}a}, {Beutler},
  {Bolton}, {Brownstein}, {Eisenstein}, {Gil-Mar{\'{\i}}n}, {Ho} \&
  {McBride}}{{Cuesta} et~al.}{2016}]{2016MNRAS.457.1770C}
{Cuesta} A.~J. et~al., 2016, MNRAS, 457, 1770, \eprintc{1509.06371},
  \adsurl{http://adsabs.harvard.edu/abs/2016MNRAS.457.1770C}

\bibitem[\protect\citeauthoryear{{Dawson}, {Kneib}, {Percival}, {Alam},
  {Albareti}, {Anderson}, {Armengaud}, {Aubourg}, {Bailey}, {Bautista},
  {Berlind}, {Bershady} \& {Beutler}}{{Dawson}
  et~al.}{2016}]{2016AJ....151...44D}
{Dawson} K.~S. et~al., 2016, AJ, 151, 44, \eprintc{1508.04473},
  \adsurl{http://adsabs.harvard.edu/abs/2016AJ....151...44D}

\bibitem[\protect\citeauthoryear{{Dawson}, {Schlegel}, {Ahn}, {Anderson},
  {Aubourg}, {Bailey}, {Barkhouser}, {Bautista}, {Beifiori} \&
  {Berlind}}{{Dawson} et~al.}{2013}]{2013AJ....145...10D}
{Dawson} K.~S. et~al., 2013, AJ, 145, 10, \eprintc{1208.0022},
  \adsurl{http://adsabs.harvard.edu/abs/2013AJ....145...10D}

\bibitem[\protect\citeauthoryear{{Doi}, {Tanaka}, {Fukugita}, {Gunn}, {Yasuda},
  {Ivezi{\'c}}, {Brinkmann}, {de Haars}, {Kleinman}, {Krzesinski} \& {French
  Leger}}{{Doi} et~al.}{2010}]{2010AJ....139.1628D}
{Doi} M. et~al., 2010, AJ, 139, 1628, \eprintc{1002.3701},
  \adsurl{http://adsabs.harvard.edu/abs/2010AJ....139.1628D}

\bibitem[\protect\citeauthoryear{{Ducout}, {Bouchet}, {Colombi}, {Pogosyan} \&
  {Prunet}}{{Ducout} et~al.}{2013}]{2013MNRAS.429.2104D}
{Ducout} A. et~al., 2013, MNRAS, 429, 2104, \eprintc{1209.1223},
  \adsurl{http://adsabs.harvard.edu/abs/2013MNRAS.429.2104D}

\bibitem[\protect\citeauthoryear{{Einasto}, {Lietzen}, {Tempel}, {Gramann},
  {Liivam{\"a}gi} \& {Einasto}}{{Einasto} et~al.}{2014}]{2014A&A...562A..87E}
{Einasto} M. et~al., 2014, A\&A, 562, A87, \eprintc{1401.3226},
  \adsurl{http://adsabs.harvard.edu/abs/2014A\%26A...562A..87E}

\bibitem[\protect\citeauthoryear{{Eisenstein}, {Weinberg}, {Agol}, {Aihara},
  {Allende Prieto}, {Anderson}, {Arns}, {Aubourg}, {Bailey}, {Balbinot} \& et
  al.}{{Eisenstein} et~al.}{2011}]{2011AJ....142...72E}
{Eisenstein} D.~J. et~al., 2011, AJ, 142, 72, \eprintc{1101.1529},
  \adsurl{http://adsabs.harvard.edu/abs/2011AJ....142...72E}

\bibitem[\protect\citeauthoryear{{Fry} \& {Peebles}}{{Fry} \&
  {Peebles}}{1978}]{1978ApJ...221...19F}
{Fry} J.~N., {Peebles} P.~J.~E. 1978, ApJ, 221, 19,
  \adsurl{http://adsabs.harvard.edu/abs/1978ApJ...221...19F}

\bibitem[\protect\citeauthoryear{{Fukugita}, {Ichikawa}, {Gunn}, {Doi},
  {Shimasaku} \& {Schneider}}{{Fukugita} et~al.}{1996}]{1996AJ....111.1748F}
{Fukugita} M. et~al., 1996, AJ, 111, 1748,
  \adsurl{http://adsabs.harvard.edu/abs/1996AJ....111.1748F}

\bibitem[\protect\citeauthoryear{{Gazta{\~n}aga}, {Cabr{\'e}}, {Castander},
  {Crocce} \& {Fosalba}}{{Gazta{\~n}aga} et~al.}{2009}]{2009MNRAS.399..801G}
{Gazta{\~n}aga} E. et~al., 2009, MNRAS, 399, 801, \eprintc{0807.2448},
  \adsurl{http://adsabs.harvard.edu/abs/2009MNRAS.399..801G}

\bibitem[\protect\citeauthoryear{{Gleser}, {Nusser}, {Ciardi} \&
  {Desjacques}}{{Gleser} et~al.}{2006}]{2006MNRAS.370.1329G}
{Gleser} L., {Nusser} A., {Ciardi} B., {Desjacques} V. 2006, MNRAS, 370, 1329,
  \eprintc{astro-ph/0602616},
  \adsurl{http://adsabs.harvard.edu/abs/2006MNRAS.370.1329G}

\bibitem[\protect\citeauthoryear{{Gunn}, {Carr}, {Rockosi}, {Sekiguchi},
  {Berry}, {Elms}, {de Haas}, {Ivezi{\'c}}, {Knapp}, {Lupton}, {Pauls} \& {et
  al.}}{{Gunn} et~al.}{1998}]{1998AJ....116.3040G}
{Gunn} J.~E. et~al., 1998, AJ, 116, 3040, \eprintc{astro-ph/9809085},
  \adsurl{http://adsabs.harvard.edu/abs/1998AJ....116.3040G}

\bibitem[\protect\citeauthoryear{{Gunn}, {Siegmund}, {Mannery}, {Owen}, {Hull},
  {Leger}, {Carey}, {Knapp}, {York} \& {et al.}}{{Gunn}
  et~al.}{2006}]{2006AJ....131.2332G}
{Gunn} J.~E. et~al., 2006, AJ, 131, 2332, \eprintc{astro-ph/0602326},
  \adsurl{http://adsabs.harvard.edu/abs/2006AJ....131.2332G}

\bibitem[\protect\citeauthoryear{Hadwiger}{Hadwiger}{1957}]{UBHD152758}
Hadwiger H., 1957, Vorlesungen \"uber Inhalt, Oberfl\"ache und Isoperimetrie.
Die Grundlehren der mathematischen Wissenschaften in Einzeldarstellungen mit
  besonderer Ber\"ucksichtigung der Anwendungsgebiete ; 93, Springer,
  \ISBN{3-540-02151-5, }

\bibitem[\protect\citeauthoryear{{Hikage} \& {Matsubara}}{{Hikage} \&
  {Matsubara}}{2012}]{2012MNRAS.425.2187H}
{Hikage} C., {Matsubara} T. 2012, MNRAS, 425, 2187, \eprintc{1207.1183},
  \adsurl{http://adsabs.harvard.edu/abs/2012MNRAS.425.2187H}

\bibitem[\protect\citeauthoryear{{Hikage}, {Schmalzing}, {Buchert}, {Suto},
  {Kayo}, {Taruya}, {Vogeley}, {Hoyle}, {Gott} III \& {Brinkmann}}{{Hikage}
  et~al.}{2003}]{2003PASJ...55..911H}
{Hikage} C. et~al., 2003, \pasj, 55, 911, \eprintc{astro-ph/0304455},
  \adsurl{http://adsabs.harvard.edu/abs/2003PASJ...55..911H}

\bibitem[\protect\citeauthoryear{{Kerscher}}{{Kerscher}}{2001}]{2001PhRvE..64e6109K}
{Kerscher} M., 2001, \pre, 64, 056109, \eprintc{astro-ph/0102153},
  \adsurl{http://adsabs.harvard.edu/abs/2001PhRvE..64e6109K}

\bibitem[\protect\citeauthoryear{{Kerscher}, {Mecke}, {Schmalzing}, {Beisbart},
  {Buchert} \& {Wagner}}{{Kerscher} et~al.}{2001}]{2001A&A...373....1K}
{Kerscher} M. et~al., 2001, A\&A, 373, 1, \eprintc{astro-ph/0101238},
  \adsurl{http://adsabs.harvard.edu/abs/2001A&A...373....1K}

\bibitem[\protect\citeauthoryear{{Kerscher}, {Schmalzing}, {Buchert} \&
  {Wagner}}{{Kerscher} et~al.}{1996}]{1996app..conf...83K}
{Kerscher} M., {Schmalzing} J., {Buchert} T., {Wagner} H. 1996, in {Weiss} A.,
  {Raffelt} G.,  {Hillebrandt} W.,  {von Feilitzsch} F.,   {Buchert} T.,  eds,
  Astro-Particle Physics {The significance of the fluctuations in the IRAS 1.2
  Jy galaxy catalogue.}.
pp 83--98, \adsurl{http://adsabs.harvard.edu/abs/1996app..conf...83K}

\bibitem[\protect\citeauthoryear{{Kerscher}, {Schmalzing}, {Buchert} \&
  {Wagner}}{{Kerscher} et~al.}{1998}]{1998A&A...333....1K}
{Kerscher} M., {Schmalzing} J., {Buchert} T., {Wagner} H. 1998, A\&A, 333, 1,
  \eprintc{astro-ph/9704028},
  \adsurl{http://adsabs.harvard.edu/abs/1998A&A...333....1K}

\bibitem[\protect\citeauthoryear{{Kerscher}, {Schmalzing}, {Retzlaff},
  {Borgani}, {Buchert}, {Gottl{\"o}ber}, {M{\"u}ller}, {Plionis} \&
  {Wagner}}{{Kerscher} et~al.}{1997}]{1997MNRAS.284...73K}
{Kerscher} M. et~al., 1997, MNRAS, 284, 73, \eprintc{astro-ph/9606133},
  \adsurl{http://adsabs.harvard.edu/abs/1997MNRAS.284...73K}

\bibitem[\protect\citeauthoryear{{Kerscher}, {Szapudi} \& {Szalay}}{{Kerscher}
  et~al.}{2000}]{2000ApJ...535L..13K}
{Kerscher} M., {Szapudi} I., {Szalay} A.~S. 2000, ApJ, 535, L13,
  \eprintc{astro-ph/9912088},
  \adsurl{http://adsabs.harvard.edu/abs/2000ApJ...535L..13K}

\bibitem[\protect\citeauthoryear{{Kitaura}, {Rodr{\'{\i}}guez-Torres},
  {Chuang}, {Zhao}, {Prada}, {Gil-Mar{\'{\i}}n}, {Guo}, {Yepes} \&
  {Klypin}}{{Kitaura} et~al.}{2016}]{2016MNRAS.456.4156K}
{Kitaura} F.-S. et~al., 2016, MNRAS, 456, 4156, \eprintc{1509.06400},
  \adsurl{http://adsabs.harvard.edu/abs/2016MNRAS.456.4156K}

\bibitem[\protect\citeauthoryear{{Kitaura}, {Yepes} \& {Prada}}{{Kitaura}
  et~al.}{2014}]{2014MNRAS.439L..21K}
{Kitaura} F.-S., {Yepes} G., {Prada} F. 2014, MNRAS, 439, L21,
  \eprintc{1307.3285},
  \adsurl{http://adsabs.harvard.edu/abs/2014MNRAS.439L..21K}

\bibitem[\protect\citeauthoryear{{Kratochvil}, {Lim}, {Wang}, {Haiman}, {May}
  \& {Huffenberger}}{{Kratochvil} et~al.}{2012}]{2012PhRvD..85j3513K}
{Kratochvil} J.~M. et~al., 2012, \prd, 85, 103513, \eprintc{1109.6334},
  \adsurl{http://adsabs.harvard.edu/abs/2012PhRvD..85j3513K}

\bibitem[\protect\citeauthoryear{{Laureijs}, {Amiaux}, {Arduini},
  {Augu{\`e}res}, {Brinchmann}, {Cole}, {Cropper}, {Dabin}, {Duvet}, {Ealet} \&
  et al.}{{Laureijs} et~al.}{2011}]{2011arXiv1110.3193L}
{Laureijs} R. et~al., 2011, ArXiv e-prints, \eprintc{1110.3193},
  \adsurl{http://adsabs.harvard.edu/abs/2011arXiv1110.3193L}

\bibitem[\protect\citeauthoryear{{Levi}, {Bebek}, {Beers}, {Blum}, {Cahn},
  {Eisenstein}, {Flaugher}, {Honscheid}, {Kron}, {Lahav}, {McDonald}, {Roe},
  {Schlegel} \& {representing the DESI collaboration}}{{Levi}
  et~al.}{2013}]{2013arXiv1308.0847L}
{Levi} M. et~al., 2013, ArXiv e-prints, \eprintc{1308.0847},
  \adsurl{http://adsabs.harvard.edu/abs/2013arXiv1308.0847L}

\bibitem[\protect\citeauthoryear{{Lupton}, {Gunn}, {Ivezi{\'c}}, {Knapp} \&
  {Kent}}{{Lupton} et~al.}{2001}]{2001ASPC..238..269L}
{Lupton} R. et~al., 2001, in {Harnden} Jr. F.~R.,  {Primini} F.~A.,   {Payne}
  H.~E.,  eds, Astronomical Data Analysis Software and Systems X Vol.~238 of
  Astronomical Society of the Pacific Conference Series, {The SDSS Imaging
  Pipelines}.
p.~269, \eprintc{astro-ph/0101420},
  \adsurl{http://adsabs.harvard.edu/abs/2001ASPC..238..269L}

\bibitem[\protect\citeauthoryear{{Mar{\'{\i}}n}}{{Mar{\'{\i}}n}}{2011}]{2011ApJ...737...97M}
{Mar{\'{\i}}n} F., 2011, ApJ, 737, 97, \eprintc{1011.4530},
  \adsurl{http://adsabs.harvard.edu/abs/2011ApJ...737...97M}

\bibitem[\protect\citeauthoryear{{Mar{\'{\i}}n}, {Blake}, {Poole}, {McBride},
  {Brough}, {Colless}, {Contreras}, {Couch}, {Croton}, {Croom}, {Davis},
  {Drinkwater} \& {Forster}}{{Mar{\'{\i}}n} et~al.}{2013}]{2013MNRAS.432.2654M}
{Mar{\'{\i}}n} F.~A. et~al., 2013, MNRAS, 432, 2654, \eprintc{1303.6644},
  \adsurl{http://adsabs.harvard.edu/abs/2013MNRAS.432.2654M}

\bibitem[\protect\citeauthoryear{Mecke \& Wagner}{Mecke \&
  Wagner}{1991}]{mecke1991euler}
Mecke K., Wagner H. 1991, Journal of Statistical Physics, 64, 843

\bibitem[\protect\citeauthoryear{{Mecke}, {Buchert} \& {Wagner}}{{Mecke}
  et~al.}{1994}]{1994A&A...288..697M}
{Mecke} K.~R., {Buchert} T., {Wagner} H. 1994, A\&A, 288, 697,
  \eprintc{astro-ph/9312028},
  \adsurl{http://adsabs.harvard.edu/abs/1994A&A...288..697M}

\bibitem[\protect\citeauthoryear{{Modest}, {R{\"a}th}, {Banday}, {Rossmanith},
  {S{\"u}tterlin}, {Basak}, {Delabrouille}, {G{\'o}rski} \& {Morfill}}{{Modest}
  et~al.}{2013}]{2013MNRAS.428..551M}
{Modest} H.~I. et~al., 2013, MNRAS, 428, 551, \eprintc{1209.5106},
  \adsurl{http://adsabs.harvard.edu/abs/2013MNRAS.428..551M}

\bibitem[\protect\citeauthoryear{{Munshi}, {Smidt}, {Cooray}, {Renzi},
  {Heavens} \& {Coles}}{{Munshi} et~al.}{2013}]{2013MNRAS.434.2830M}
{Munshi} D. et~al., 2013, MNRAS, 434, 2830, \eprintc{1011.5224},
  \adsurl{http://adsabs.harvard.edu/abs/2013MNRAS.434.2830M}

\bibitem[\protect\citeauthoryear{{Nakagami}, {Matsubara}, {Schmalzing} \&
  {Jing}}{{Nakagami} et~al.}{2004}]{2004astro.ph..8428N}
{Nakagami} T., {Matsubara} T., {Schmalzing} J., {Jing} Y. 2004, ArXiv
  Astrophysics e-prints, \eprintc{astro-ph/0408428},
  \adsurl{http://adsabs.harvard.edu/abs/2004astro.ph..8428N}

\bibitem[\protect\citeauthoryear{{Neyrinck}, {Szapudi} \& {Szalay}}{{Neyrinck}
  et~al.}{2009}]{2009ApJ...698L..90N}
{Neyrinck} M.~C., {Szapudi} I., {Szalay} A.~S. 2009, ApJ, 698, L90,
  \eprintc{0903.4693},
  \adsurl{http://adsabs.harvard.edu/abs/2009ApJ...698L..90N}

\bibitem[\protect\citeauthoryear{{Neyrinck}, {Wang}, {Falck}, {Szapudi} \&
  {Szalay}}{{Neyrinck} et~al.}{2011}]{2011AAS...21823304N}
{Neyrinck} M.~C. et~al., 2011, in American Astronomical Society Meeting
  Abstracts \#218 {The Log-Density as a Better Cosmological Density Variable}.
p. \#233.04, \adsurl{http://adsabs.harvard.edu/abs/2011AAS...21823304N}

\bibitem[\protect\citeauthoryear{{Padmanabhan}, {Schlegel}, {Finkbeiner},
  {Barentine}, {Blanton}, {Brewington}, {Gunn}, {Harvanek}, {Hogg} \& {et
  al.}}{{Padmanabhan} et~al.}{2008}]{2008ApJ...674.1217P}
{Padmanabhan} N. et~al., 2008, ApJ, 674, 1217, \eprintc{astro-ph/0703454},
  \adsurl{http://adsabs.harvard.edu/abs/2008ApJ...674.1217P}

\bibitem[\protect\citeauthoryear{{Peebles} \& {Groth}}{{Peebles} \&
  {Groth}}{1975}]{1975ApJ...196....1P}
{Peebles} P.~J.~E., {Groth} E.~J. 1975, ApJ, 196, 1,
  \adsurl{http://adsabs.harvard.edu/abs/1975ApJ...196....1P}

\bibitem[\protect\citeauthoryear{{Percival}, {Ross}, {S{\'a}nchez}, {Samushia},
  {Burden}, {Crittenden}, {Cuesta}, {Magana}, {Manera} \& {Beutler}}{{Percival}
  et~al.}{2014}]{2014MNRAS.439.2531P}
{Percival} W.~J. et~al., 2014, MNRAS, 439, 2531, \eprintc{1312.4841},
  \adsurl{http://adsabs.harvard.edu/abs/2014MNRAS.439.2531P}

\bibitem[\protect\citeauthoryear{{Petri}, {Haiman}, {Hui}, {May} \&
  {Kratochvil}}{{Petri} et~al.}{2013}]{2013PhRvD..88l3002P}
{Petri} A. et~al., 2013, \prd, 88, 123002, \eprintc{1309.4460},
  \adsurl{http://adsabs.harvard.edu/abs/2013PhRvD..88l3002P}

\bibitem[\protect\citeauthoryear{{Pier}, {Munn}, {Hindsley}, {Hennessy},
  {Kent}, {Lupton} \& {Ivezi{\'c}}}{{Pier} et~al.}{2003}]{2003AJ....125.1559P}
{Pier} J.~R. et~al., 2003, AJ, 125, 1559, \eprintc{astro-ph/0211375},
  \adsurl{http://adsabs.harvard.edu/abs/2003AJ....125.1559P}

\bibitem[\protect\citeauthoryear{{Planck Collaboration}, {Ade}, {Aghanim},
  {Akrami}, {Aluri}, {Arnaud}, {Ashdown}, {Aumont}, {Baccigalupi}, {Banday} \&
  et al.}{{Planck Collaboration XVI}}{2016}]{2016A&A...594A..16P}
{Planck Collaboration XVI}, 2016, A\&A, 594, A16, \eprintc{1506.07135},
  \adsurl{http://adsabs.harvard.edu/abs/2016A\%26A...594A..16P}

\bibitem[\protect\citeauthoryear{{Planck Collaboration}, {Ade}, {Aghanim},
  {Armitage-Caplan}, {Arnaud}, {Ashdown}, {Atrio-Barandela}, {Aumont},
  {Baccigalupi}, {Banday} \& et al.}{{Planck Collaboration XXIII}}{2014a}]{2014A&A...571A..23P}
{Planck Collaboration XXIII}, 2014a, A\&A, 571, A23, \eprintc{1303.5083},
  \adsurl{http://adsabs.harvard.edu/abs/2014A\%26A...571A..23P}

\bibitem[\protect\citeauthoryear{{Planck Collaboration}, {Ade}, {Aghanim},
  {Armitage-Caplan}, {Arnaud}, {Ashdown}, {Atrio-Barandela}, {Aumont},
  {Baccigalupi}, {Banday} \& et al.}{{Planck Collaboration XXV}}{2014b}]{2014A&A...571A..25P}
{Planck Collaboration XXV}, 2014b, A\&A, 571, A25, \eprintc{1303.5085},
  \adsurl{http://adsabs.harvard.edu/abs/2014A\%26A...571A..25P}

\bibitem[\protect\citeauthoryear{{Platz{\"o}der} \& {Buchert}}{{Platz{\"o}der}
  \& {Buchert}}{1996}]{1996app..conf..251P}
{Platz{\"o}der} M., {Buchert} T. 1996, in {Weiss} A.,  {Raffelt} G.,
  {Hillebrandt} W.,  {von Feilitzsch} F.,   {Buchert} T.,  eds, Astro-Particle
  Physics {Applications of Minkowski-Functionals to the Statistical Analysis of
  Dark Matter Models}.
p.~251, \eprintc{astro-ph/9509014},
  \adsurl{http://adsabs.harvard.edu/abs/1996app..conf..251P}

\bibitem[\protect\citeauthoryear{{Pratten} \& {Munshi}}{{Pratten} \&
  {Munshi}}{2012}]{2012MNRAS.423.3209P}
{Pratten} G., {Munshi} D. 2012, MNRAS, 423, 3209, \eprintc{1108.1985},
  \adsurl{http://adsabs.harvard.edu/abs/2012MNRAS.423.3209P}

\bibitem[\protect\citeauthoryear{{Reid}, {Ho}, {Padmanabhan}, {Percival},
  {Tinker}, {Tojeiro}, {White}, {Eisenstein}, {Maraston}, {Ross} \&
  {S{\'a}nchez}}{{Reid} et~al.}{2016}]{2016MNRAS.455.1553R}
{Reid} B. et~al., 2016, MNRAS, 455, 1553, \eprintc{1509.06529},
  \adsurl{http://adsabs.harvard.edu/abs/2016MNRAS.455.1553R}

\bibitem[\protect\citeauthoryear{{Sahni}, {Sathyaprakash} \&
  {Shandarin}}{{Sahni} et~al.}{1998}]{1998ApJ...495L...5S}
{Sahni} V., {Sathyaprakash} B.~S., {Shandarin} S.~F. 1998, ApJ, 495, L5,
  \eprintc{astro-ph/9801053},
  \adsurl{http://adsabs.harvard.edu/abs/1998ApJ...495L...5S}

\bibitem[\protect\citeauthoryear{{Sathyaprakash}, {Sahni} \&
  {Shandarin}}{{Sathyaprakash} et~al.}{1998}]{1998ApJ...508..551S}
{Sathyaprakash} B.~S., {Sahni} V., {Shandarin} S. 1998, ApJ, 508, 551,
  \eprintc{astro-ph/9805285},
  \adsurl{http://adsabs.harvard.edu/abs/1998ApJ...508..551S}

\bibitem[\protect\citeauthoryear{Schmalzing}{Schmalzing}{1999}]{phdSchmalzing}
Schmalzing J., 1999, PhD thesis, LMU M{\"u}nchen

\bibitem[\protect\citeauthoryear{{Schmalzing} \& {Buchert}}{{Schmalzing} \&
  {Buchert}}{1997}]{1997ApJ...482L...1S}
{Schmalzing} J., {Buchert} T. 1997, ApJ, 482, L1, \eprintc{astro-ph/9702130},
  \adsurl{http://adsabs.harvard.edu/abs/1997ApJ...482L...1S}

\bibitem[\protect\citeauthoryear{{Schmalzing}, {Buchert}, {Melott}, {Sahni},
  {Sathyaprakash} \& {Shandarin}}{{Schmalzing}
  et~al.}{1999}]{1999ApJ...526..568S}
{Schmalzing} J. et~al., 1999, ApJ, 526, 568, \eprintc{astro-ph/9904384},
  \adsurl{http://adsabs.harvard.edu/abs/1999ApJ...526..568S}

\bibitem[\protect\citeauthoryear{{Schmalzing} \& {Gorski}}{{Schmalzing} \&
  {Gorski}}{1998}]{1998MNRAS.297..355S}
{Schmalzing} J., {Gorski} K.~M. 1998, MNRAS, 297, 355,
  \eprintc{astro-ph/9710185},
  \adsurl{http://adsabs.harvard.edu/abs/1998MNRAS.297..355S}

\bibitem[\protect\citeauthoryear{{Schmalzing}, {Gottl{\"o}ber}, {Klypin} \&
  {Kravtsov}}{{Schmalzing} et~al.}{1999}]{1999MNRAS.309.1007S}
{Schmalzing} J., {Gottl{\"o}ber} S., {Klypin} A.~A., {Kravtsov} A.~V. 1999,
  MNRAS, 309, 1007, \eprintc{astro-ph/9906475},
  \adsurl{http://adsabs.harvard.edu/abs/1999MNRAS.309.1007S}

\bibitem[\protect\citeauthoryear{{Schmalzing}, {Kerscher} \&
  {Buchert}}{{Schmalzing} et~al.}{1996}]{1996dmu..conf..281S}
{Schmalzing} J., {Kerscher} M., {Buchert} T. 1996, in {Bonometto} S.,
  {Primack} J.~R.,   {Provenzale} A.,  eds, Dark Matter in the Universe
  {Minkowski Functionals in Cosmology}.
p.~281, \eprintc{astro-ph/9508154},
  \adsurl{http://adsabs.harvard.edu/abs/1996dmu..conf..281S}

\bibitem[\protect\citeauthoryear{{Schmalzing}, {Takada} \&
  {Futamase}}{{Schmalzing} et~al.}{2000}]{2000ApJ...544L..83S}
{Schmalzing} J., {Takada} M., {Futamase} T. 2000, ApJ, 544, L83,
  \adsurl{http://adsabs.harvard.edu/abs/2000ApJ...544L..83S}

\bibitem[\protect\citeauthoryear{{Slepian} \& {Eisenstein}}{{Slepian} \&
  {Eisenstein}}{2015}]{2015MNRAS.448....9S}
{Slepian} Z., {Eisenstein} D.~J. 2015, MNRAS, 448, 9, \eprintc{1411.4052},
  \adsurl{http://adsabs.harvard.edu/abs/2015MNRAS.448....9S}

\bibitem[\protect\citeauthoryear{{Slepian} \& {Eisenstein}}{{Slepian} \&
  {Eisenstein}}{2016}]{2016arXiv160703109S}
{Slepian} Z., {Eisenstein} D.~J. 2016, ArXiv e-prints, \eprintc{1607.03109},
  \adsurl{http://adsabs.harvard.edu/abs/2016arXiv160703109S}

\bibitem[\protect\citeauthoryear{{Slepian}, {Eisenstein}, {Beutler}, {Cuesta},
  {Ge}, {Gil-Mar{\'{\i}}n}, {Ho}, {Kitaura} \& {McBride}}{{Slepian}
  et~al.}{2015}]{2015arXiv151202231S}
{Slepian} Z. et~al., 2015, ArXiv e-prints, \eprintc{1512.02231},
  \adsurl{http://adsabs.harvard.edu/abs/2015arXiv151202231S}

\bibitem[\protect\citeauthoryear{{Smee}, {Gunn}, {Uomoto}, {Roe}, {Schlegel},
  {Rockosi}, {Carr}, {Leger}, {Dawson}, {Olmstead} \& {et al.}}{{Smee}
  et~al.}{2013}]{2013AJ....146...32S}
{Smee} S.~A. et~al., 2013, AJ, 146, 32, \eprintc{1208.2233},
  \adsurl{http://adsabs.harvard.edu/abs/2013AJ....146...32S}

\bibitem[\protect\citeauthoryear{{Smith}, {Tucker}, {Kent}, {Richmond},
  {Fukugita}, {Ichikawa}, {Ichikawa}, {Jorgensen}, {Uomoto}, {Gunn} \& {et
  al.}}{{Smith} et~al.}{2002}]{2002AJ....123.2121S}
{Smith} J.~A. et~al., 2002, AJ, 123, 2121, \eprintc{astro-ph/0201143},
  \adsurl{http://adsabs.harvard.edu/abs/2002AJ....123.2121S}

\bibitem[\protect\citeauthoryear{Stratonovich}{Stratonovich}{1963}]{Stratonovich1963}
Stratonovich R.~L., 1963, Topics in the theory of random noise.
Vol. Vol. 1, Gordon and Breach, New York

\bibitem[\protect\citeauthoryear{{Wang}, {Neyrinck}, {Szapudi}, {Szalay},
  {Chen}, {Lesgourgues}, {Riotto} \& {Sloth}}{{Wang}
  et~al.}{2011}]{2011ApJ...735...32W}
{Wang} X. et~al., 2011, ApJ, 735, 32, \eprintc{1103.2166},
  \adsurl{http://cdsads.u-strasbg.fr/abs/2011ApJ...735...32W}

\bibitem[\protect\citeauthoryear{{White}, {Tinker} \& {McBride}}{{White}
  et~al.}{2014}]{2014MNRAS.437.2594W}
{White} M., {Tinker} J.~L., {McBride} C.~K. 2014, MNRAS, 437, 2594,
  \eprintc{1309.5532},
  \adsurl{http://adsabs.harvard.edu/abs/2014MNRAS.437.2594W}

\bibitem[\protect\citeauthoryear{{White}}{{White}}{1979}]{1979MNRAS.186..145W}
{White} S.~D.~M., 1979, MNRAS, 186, 145,
  \adsurl{http://adsabs.harvard.edu/abs/1979MNRAS.186..145W}

\bibitem[\protect\citeauthoryear{{Wiegand}, {Buchert} \& {Ostermann}}{{Wiegand}
  et~al.}{2014}]{2014MNRAS.443..241W}
{Wiegand} A., {Buchert} T., {Ostermann} M. 2014, MNRAS, 443, 241,
  \eprintc{1311.3661},
  \adsurl{http://adsabs.harvard.edu/abs/2014MNRAS.443..241W}

\bibitem[\protect\citeauthoryear{{York}, {Adelman}, {Anderson} Jr., {Anderson},
  {Annis}, {Bahcall}, {Bakken}, {Barkhouser}, {Bastian} \& {SDSS
  Collaboration}}{{York} et~al.}{2000}]{2000AJ....120.1579Y}
{York} D.~G. et~al., 2000, AJ, 120, 1579, \eprintc{astro-ph/0006396},
  \adsurl{http://adsabs.harvard.edu/abs/2000AJ....120.1579Y}

\bibitem[\protect\citeauthoryear{{Yoshiura}, {Shimabukuro}, {Takahashi} \&
  {Matsubara}}{{Yoshiura} et~al.}{2016}]{2016arXiv160202351Y}
{Yoshiura} S., {Shimabukuro} H., {Takahashi} K., {Matsubara} T. 2016, ArXiv
  e-prints, \eprintc{1602.02351},
  \adsurl{http://adsabs.harvard.edu/abs/2016arXiv160202351Y}

\end{thebibliography}

\bsp
\label{lastpage}
\end{document}